\PassOptionsToPackage{hyphens}{url}

\documentclass[acmsmall]{acmart}
\acmJournal{TRETS}
\usepackage{amsmath,amssymb,amsfonts}
\usepackage{algorithmic}
\usepackage[ruled, noline, noend, linesnumbered]{algorithm2e}
\SetCommentSty{font=regular}

\let\oldnl\nl
\newcommand{\nonl}{\renewcommand{\nl}{\let\nl\oldnl}}

\usepackage{subcaption}
\usepackage{rotating}
\usepackage{graphicx}
\usepackage{textcomp}
\usepackage{pgfplots}
\pgfplotsset{compat=1.18}
\usepackage{threeparttable, booktabs}
\usepackage{rotating}
\usepackage{multirow}
\usepackage{xcolor}
\usepackage{xspace}
\usepackage{hyperref}
\hypersetup{colorlinks=true, urlcolor=blue}

\definecolor{codegreen}{rgb}{0,0.6,0}
\definecolor{codegray}{rgb}{0.5,0.5,0.5}
\definecolor{codepurple}{rgb}{0.58,0,0.82}
\definecolor{backcolour}{rgb}{0.95,0.95,0.92}

\newcommand{\ryan}[1]{\textcolor{red}{}}
\newcommand{\aba}[1]{\textcolor{blue}{}}
\newcommand{\zhenghua}[1]{\textcolor{orange}{}}
\newcommand{\francesco}[1]{\textcolor{green}{}}

\newcommand{\revision}[1]{#1}

\newcommand{\codename}{{\sc cgra4ml}\xspace}
\newcommand{\hlsml}{{\sc hls4ml}\xspace}
\newcommand{\qkeras}{{\sc qkeras}\xspace}
\newcommand{\finn}{{\sc finn}\xspace}

\title[\codename: A Framework for Scientific Edge ML]{\codename: A Hardware/Software Framework to Implement Neural Networks for Scientific Edge Computing}

\author{G. Abarajithan}
\orcid{0000-0001-9768-5349}
\affiliation{%
  \department{Department of Computer Science and Engineering}
  \institution{University of California, San Diego}
  \city{La Jolla}
  \state{CA}
  \country{USA}
}
\email{agnaneswaran@ucsd.edu}

\author{Zhenghua Ma}
\orcid{0009-0002-1046-5883}
\affiliation{%
  \department{Department of Computer Science and Engineering}
  \institution{University of California, San Diego}
  \city{La Jolla}
  \state{CA}
  \country{USA}
}
\email{zhm007@ucsd.edu}

\author{Ravidu Munasinghe}
\orcid{0009-0009-4544-9542}
\affiliation{%
  \department{Department of Electronic and Telecommunication}
  \institution{University of Moratuwa}
  \city{Moratuwa}
  \state{Western Province}
  \country{Sri Lanka}
}
\email{raviduhm@gmail.com}

\author{Francesco Restuccia}
\orcid{0000-0001-6955-1888}
\affiliation{%
  \department{Department of Computer Science and Engineering}
  \institution{University of California, San Diego}
  \city{La Jolla}
  \state{CA}
  \country{USA}
}
\email{frestuccia@ucsd.edu}

\author{Ryan Kastner}
\orcid{0000-0001-9062-5570}
\affiliation{%
  \department{Department of Computer Science and Engineering}
  \institution{University of California, San Diego}
  \city{La Jolla}
  \state{CA}
  \country{USA}
}
\email{kastner@ucsd.edu}

\keywords{FPGA, reconfigurable computing, edge computing, machine learning, CGRA}

\ccsdesc[500]{Hardware~Reconfigurable logic applications}
\ccsdesc[300]{Computing methodologies~Machine learning}

\acmJournal{TRETS}
\acmYear{2026}
\acmMonth{3}
\acmDOI{10.1145/3801097}

\begin{document}

\begin{abstract}

The scientific community increasingly relies on machine learning (ML) for near-sensor processing, leveraging its strengths in tasks such as pattern recognition, anomaly detection, and real-time decision-making. 
These deployments demand accelerators that combine extremely high performance with programmability, ease of integration, and straightforward verification. 
We present \codename, an open-source, modular framework that generates parameterizable CGRA accelerators in synthesizable SystemVerilog RTL, tailored to common ML compute patterns found in scientific applications. 
The framework supports seamless system integration through AXI-compliant interfaces and open-source DMA components, and it includes automatic firmware generation for programming the accelerator. 
A comprehensive verification suite and a runtime firmware stack further support deployment across diverse SoC platforms. 
\codename provides a modular, full-stack infrastructure, including a Python API, SystemVerilog hardware, TCL toolflows, and a C runtime, which facilitates easy integration and experimentation, allowing scientists to focus on innovation rather than dealing with the intricacies of hardware design and optimization.
We demonstrate the effectiveness of \codename to implement common scientific edge neural networks using ASIC and FPGA design flows. 



\end{abstract}

\maketitle

Repository: \href{https://github.com/KastnerRG/cgra4ml}{github.com/KastnerRG/cgra4ml}

\section{Introduction}

Modern scientific discovery increasingly relies on machine learning to process vast volumes of data at the edge, where sensors capture events in real-time and decisions must be made within microseconds~\cite{duarte2018fast}. 
From high-energy physics experiments and remote sensing arrays to in-situ biomedical diagnostics and fusion plasma control, these scientific applications demand neural network inference on custom hardware architectures to achieve the required inference speed and energy efficiency with extremely high throughput, low latencies, and tight energy budgets~\cite{autoqkeras}. 
Such custom hardware designs must be closely integrated with host microprocessors and memory subsystems to achieve the highest performance.
At the same time, these workloads increasingly utilize complex, evolving neural network models, necessitating a degree of programmability and architectural flexibility. 
Previous scientific edge computing frameworks such as \hlsml~\cite{hls4ml} and \finn~\cite{umuroglu2017finn} have enabled small neural networks on FPGAs through high-level synthesis. 
However, they struggle to scale to deeper models, offer limited flexibility in hardware reuse, and lack production-ready verification and integration flows.
Therefore, the scientific computing community is in need of a framework that supports larger models with a user-friendly interface, support for various quantization levels, modular hardware architecture, production-ready firmware generation, complete system integration, and verification.



To address these requirements, we present \codename, an open-source framework for deploying modern neural networks in scientific edge computing environments. 
\codename generates parametrizable, reusable Coarse-Grained Reconfigurable Array (CGRA) hardware engines and associated runtime systems, enabling scalable inference of complex models across FPGAs and ASICs. 
The framework delivers a holistic solution, comprising a Python-based frontend for quantized model specification, a SystemVerilog RTL backend for the CGRA, a modular C runtime firmware, and an integrated verification suite. 

\codename is targeted to two user groups:
(1) Scientists, embedded / IoT researchers who want to build and implement modern neural networks that cannot be deployed using current popular frameworks for scientific computing due to their size and complexity. 
(2) Hardware system designers who want to iteratively test new optimizations such as new datatypes, dataflow, and compression techniques.

The major contributions of \codename include:

\begin{itemize}
    \item A hardware + software + firmware co-design framework to specify, train, quantize, and implement diverse neural networks on FPGA and ASIC as CGRA-powered SoCs.
    \item An open-source CGRA architecture with dynamic reconfiguration, unified dataflow, and an efficient PE design to maximize hardware utilization for modern neural networks.
    \item RTL-based high-performance, modular SoC design with open-source DMAs and AXI-compliant interfaces to support vendor-agnostic FPGA and ASIC design flows.
    \item Automated firmware generation enabling hardware/software partitioning that easily integrates the CGRA with on-chip processors. 
    \item End-to-end implementation and test on AMD Xilinx FPGAs and ASIC flow using Cadence tools to be taped out as an SoC with ARM-based NanoSoC platform.
    \item Extensive verification support to test the neural network model against the generated RTL design and C runtime through randomized transactional testbenches.
\end{itemize}

We evaluate \codename on representative scientific workloads, including ResNet-50 for high-resolution image classification and PointNet for particle cloud inference. 
Results on AMD-Xilinx FPGAs and Cadence ASIC flows demonstrate that \codename achieves competitive performance-per-watt, extensibility, and scalability, making it a robust and practical solution for real-time scientific inference at the edge.

\section{Background and Related Work}
\label{sec:background}

This section outlines the requirements for a hardware-accelerated system development in scientific computing.
We first introduce the need for neural networks in scientific computing and practical use cases.
We then discuss two ways of implementing such networks: a layer-by-layer pipelined implementation where each layer becomes a pipeline stage in hardware and the accelerator-based implementation where all layers reuse the same reconfigurable hardware.
\hlsml~\cite{hls4ml} and \finn~\cite{umuroglu2017finn} take the first approach, while several accelerators~\cite{surveycnndnn2023}~\cite{surveyefficienthw2022} and frameworks such as Apache TVM with VTA accelerator~\cite{vta} take the second.
We then outline the uniqueness and utility of a CGRA-type accelerator architecture to solve this problem and discuss related research in this area.
Consequently, we summarize the needs of the scientific computing community that are not addressed by other generic frameworks as motivation for \codename.

\begin{figure}
   \centering
   \includegraphics[width=0.8\linewidth]{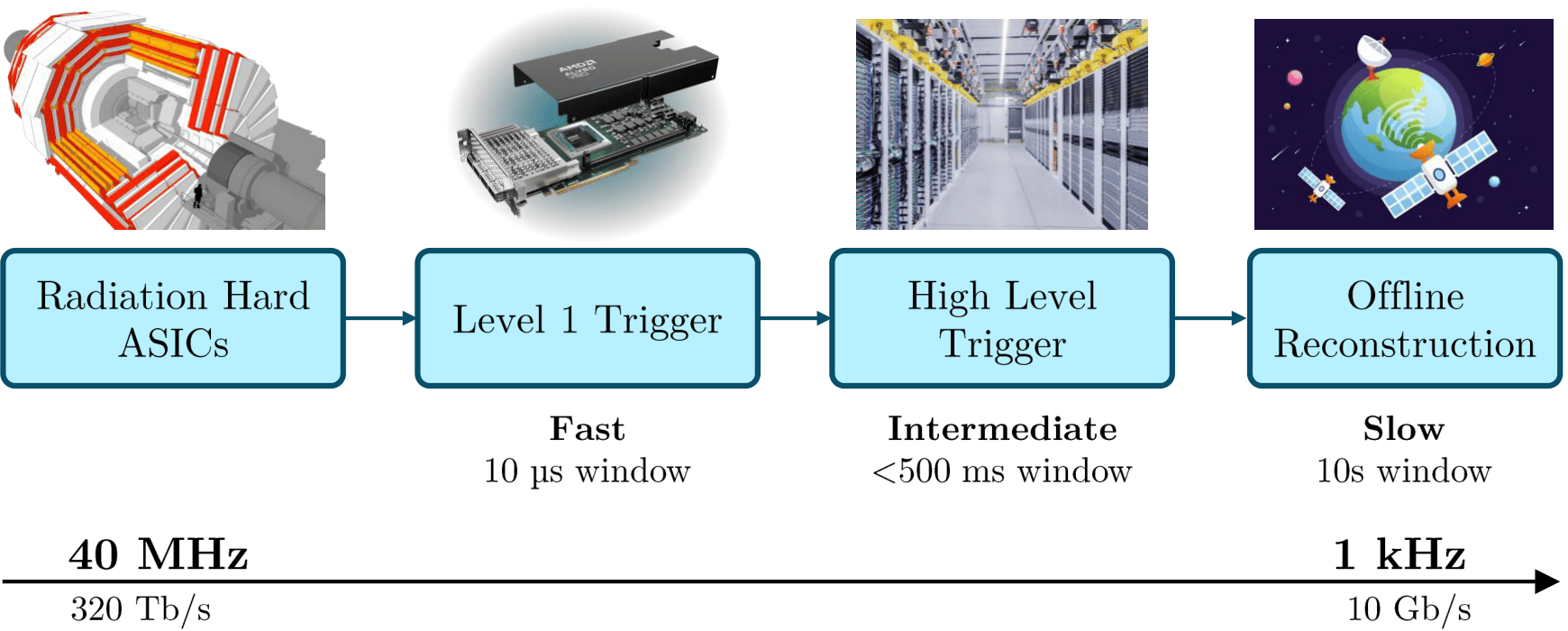}
   \caption{
   Data processing pipeline at Large Hadron Collider, CERN. The detectors at LHC produce data at a rate of 320 Tb/s during collisions. Due to the infeasibility of transferring data at such a high rate, the events are filtered through a multi-step pipeline. Small neural networks like autoencoders are implemented using \hlsml on radiation-hardened ASICs. More sophisticated models are implemented on FPGAs (L1) and servers (L2) to filter the data fully. With \codename enabling larger models at low power, more processing can be moved to the edge (towards the left), resulting in more robust filtering.
   }
   \Description{Data processing pipeline at Large Hadron Collider, CERN. }
   \label{fig:lhc}
\end{figure}

\subsection{Neural Networks for Scientific Computing}

Machine learning is fueling scientific discoveries in particle physics, materials, dark matter, cosmology, nuclear physics, biomedical engineering, and health monitoring~\cite{deiana2022applications}.
Specific examples include bio-signal classification~\cite{ahmed2024}, tracking magnetohydrodynamic (MHD) instabilities in fusion reactors~\cite{wei2023fpga}, reinforcement learning for accelerator beam control~\cite{duarte2022fastml}, pulse detection for anti-neutrino detection~\cite{axani2024rfsoc}, and accelerator controls for beam loss deblending~\cite{arnold2023edge}. 
Scientific computing increasingly relies on neural networks to process data with requirements for extremely low latencies and high throughputs. 

The Large Hadron Collider (LHC) has $\mathcal{O}(100M)$ individual sensors generating data for each of the 40 million proton beam collisions per second~\cite{part2022}.  
LHC trigger systems filter events with sub-microsecond latency and 40 MHz throughput requirements using FPGAs~\cite{duarte2018fast} as shown in Fig. \ref{fig:lhc}.
In each level, data is compressed, analyzed, and filtered.
While traditional algorithms have been used for data filtering in the past, neural networks accelerated in FPGA and ASICs are replacing them.
For example, neural network-based autoencoders are implemented as ASICs in the L0 trigger to compress and move data out of the sensors~\cite{di2021reconfigurable}.
L1 uses more complex networks accelerated on FPGAs, and L2 triggers and offline reconstruction use even bigger models executed on GPUs.
User-friendly frameworks for hardware implementation have fueled the wide adoption of small neural networks in lower-level triggers.
\hlsml is one of the more popular among scientific edge accelerator frameworks.

\revision{
Large models, such as Particle Transformer~\cite{part2022} for particle jet classification, are being explored by the physics community. 
However, for low-latency FPGA trigger use, they have not yet shown compelling gains, so practical deployments remain GPU-based. 
This is partly a chicken-and-egg issue.
Since \hlsml implementation of models requires more and more resources with increasing model size, almost all the effort from the scientific computing community goes into making their models smaller to fit within an FPGA.
 Multi-FPGA routes, such as AIgean~\cite{AIgean}, point to the possibility of using larger models but are comparatively cumbersome for trigger pipelines.
}

\subsection{Dataflow-style NN Implementation: \hlsml, \finn}

\begin{figure}
    \centering
    \includegraphics[width=0.7\linewidth]{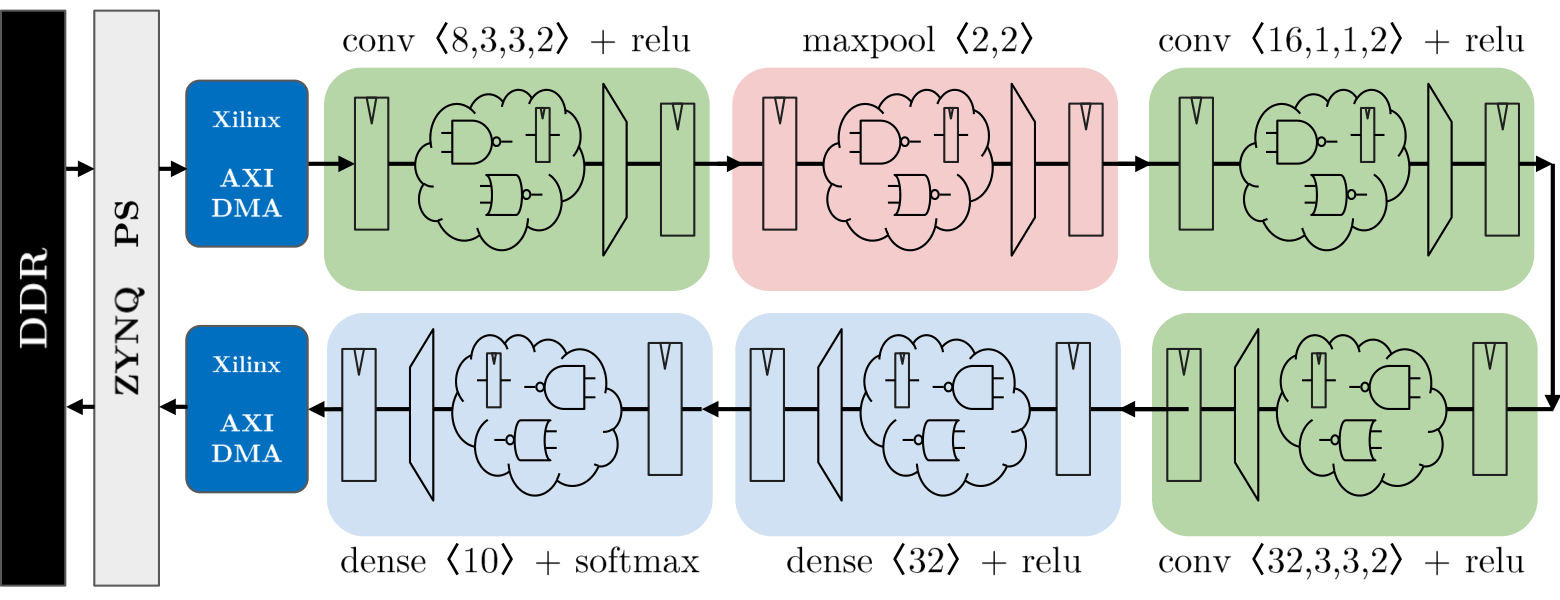}
    \caption{Layer-by-layer dataflow-style implementation of neural networks using \hlsml. \hlsml offers multiple backends, one per family of devices. Each backend is a collection of layers defined as templated HLS. Each neural network layer is implemented as a separate datapath. Xilinx DMA IPs are connected to the HLS4ML design to move data in and out. \finn has a similar dataflow-style implementation. In contrast, \codename reuses the same CGRA multiple times to process a layer.}
    \Description{Layer-by-layer spatial implementation of neural networks using \hlsml}
    \label{fig:hls4ml}
\end{figure}

\hlsml is a Python library developed for low-latency, high-throughput edge machine learning inference~\cite{hls4ml}. 
\hlsml primarily supports models built and trained using QKeras, a library from Google for low-precision machine learning~\cite{qkeras}. 
\hlsml converts each layer of the model into a customized high-level synthesis (HLS) hardware block and connects them together to form an IP for the entire model (see Fig.~\ref{fig:hls4ml}). 
The user controls \emph{reuse factor} per layer, which defines the number of times a multiplier/accumulator is reused. 
The reuse factor provides a way to control a layer's resource usage and performance.
%
Balancing performance and resource utilization is highly dependent on the specific target FPGA and involves iterative manual tuning of reuse factors on a per-layer basis. This optimization effort is non-trivial and takes a significant time and effort.




Although the layer-by-layer approach may offer advantages in latency minimization, the architecture suffers from scaling issues as the size of its target NN increases. Because each layer is implemented with a unique datapath, larger NNs with tens to hundreds of layers quickly become too complex to implement on even the largest of hardware devices.
%

As the name implies, \hlsml requires a high-level synthesis (HLS) tool flow  to generate the hardware accelerator. 
\hlsml has different HLS backends (AMD Xilinx FPGA, Altera FPGA, Siemens Catapult, etc.) with different levels of support. 
The most common target is AMD Xilinx Vitis/Vivado HLS, though other backends exist in various forms (Altera HLS Compiler, Siemens Catapult HLS, etc.). 
The dependence on HLS restricts the user to the supported vendors and requires HLS tool-specific reimplementation of each hardware layer, complicating the maintenance and verification.
\hlsml's support for ASIC implementation is primarily through Catapult HLS, which currently only supports a limited subset of layers.

A common \hlsml design flow requires scientists to generate an IP for the neural network, and then engineers write firmware, perform hardware-software integration, verification, and debugging, which is an onerous task.
There are efforts to streamline this process for the scientists.
For example, \hlsml provides a PYNQ overlay, a Python-based runtime for AMD Xilinx Zynq FPGAs. 
This workflow allows the physicists to focus on building and training low-precision models for their custom applications without spending months on hardware design, implementation, and verification. 
Yet, by and large, the ML IP core integration for production-ready designs remains a separate process from \hlsml.

\finn \cite{umuroglu2017finn} is another popular open-source FPGA accelerator framework that uses a similar methodology.
\finn solely targets FPGAs though its general methodology could be ported for other hardware backends.
\finn shares many of the benefits and drawbacks of \hlsml, while being especially optimized for extremely low bitwidths, such as binary neural networks~\cite{borras2022open}.
The layer-by-layer approach is ideal for minimizing latency, since it eliminates the need to transfer partial outputs to/from the off-chip memory.
It also makes the NN implementation extremely modular, allowing any layer to be added in any order.
Yet, \finn is primarily focused on generating an IP core that must be separately integrated into a larger system.

\subsection{DNN Accelerators}

\begin{figure}
    \centering
    \includegraphics[width=0.8\linewidth]{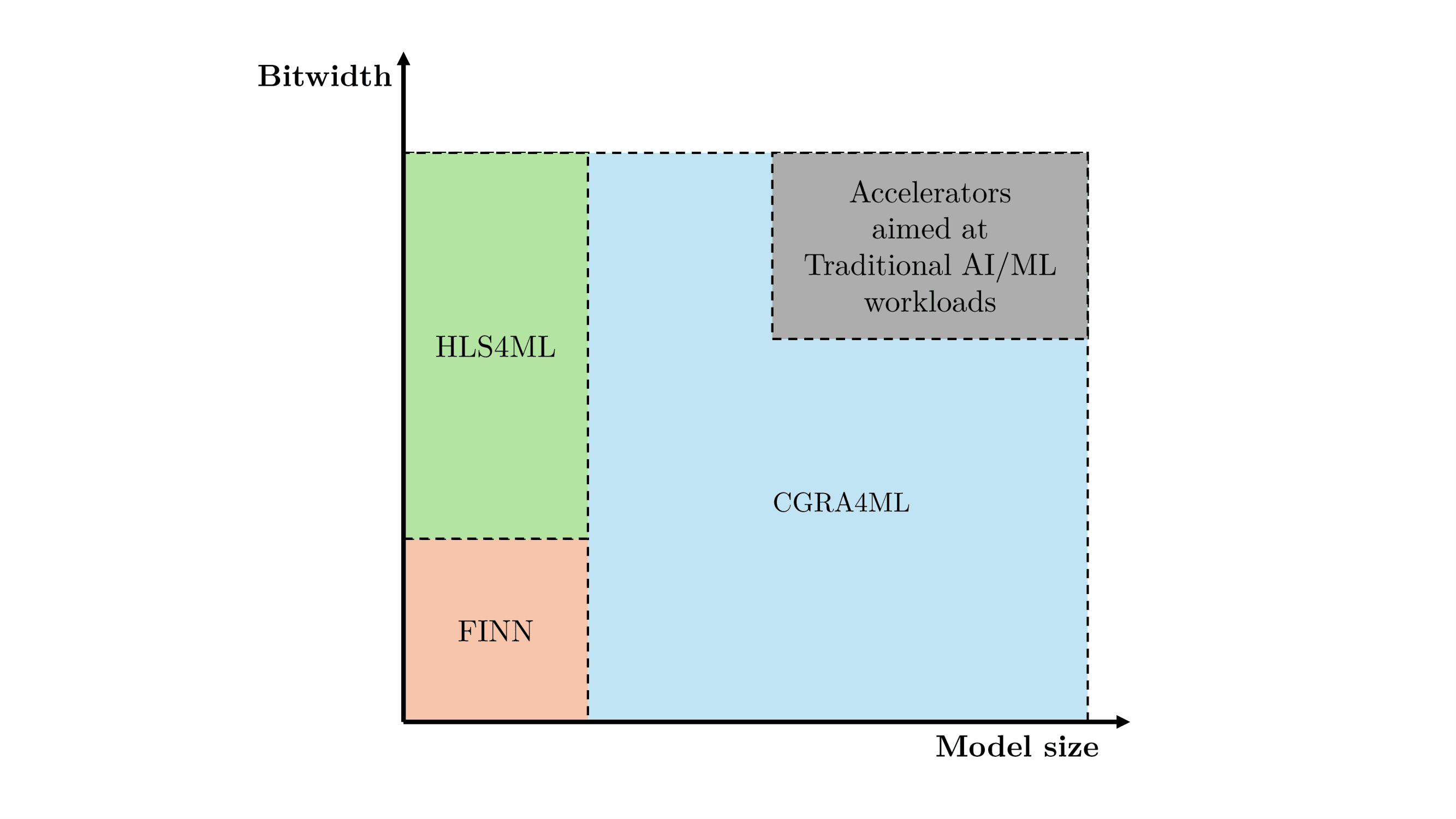}
    \vspace{-2em}
    \caption{Positioning CGRA4ML in the space of ML-to-FPGA frameworks from the perspective of the scientific computing community. \hlsml ~\cite{hls4ml} is the most popular tool in the scientific computing community, as it is user-friendly and supports the dataflow-style implementation of models of varying bitwidths and layers to an extent. \finn ~\cite{umuroglu2017finn} implements models in a similar dataflow fashion, but excels at very low bitwidths. Traditional AI accelerators are suitable for large models with 8+ bit quantization. \codename aims to fill the gap by making models with sub 8-bit quantization that are too big to implement with \hlsml and \finn.}
    \Description{Layer-by-layer spatial implementation of neural networks using \hlsml}
    \label{fig:position}
\end{figure}

Hundreds of accelerator architectures have been designed for deep neural network inference in the past decade~\cite{surveytaxo2022, surveycnndnn2023, surveyefficienthw2022}. 
This has been the default approach for the implementation of DNNs on hardware.
These accelerators can be categorized by their data reuse pattern as weight, input, output and row stationary and by their runtime flexibility into systolic arrays, CGRAs, and microcode processors.

Caffeine~\cite{zhang2023caffeine} offers an HLS systolic array to process CNNs from the Caffe ML framework using weight major and input major mappings. 
Eyeriss~\cite{eyeriss} is an energy-efficient ASIC designed to accelerate CNNs using a 2D array of 168 fairly complex processing elements, each with a 16-bit multiply-accumulate, scratchpad, and a controller. 
ShiDianNao~\cite{du2015shidiannao} uses a 2D mesh of functional units optimized towards the 2D feature maps of convolutional layers. 
Each processing element executes multiplications, additions, and comparisons using 16-bit fixed-point arithmetic. 
The kernel elements are shifted right to left and up to down and accumulate locally. 

While these architectures demonstrate strong performance and efficient data reuse, they often lack the broader system-level features required for scientific edge applications. 
Many of these accelerators are optimized for fixed model types, offering limited programmability and making it challenging to re-target them for newer or more diverse neural networks. 
Integration into embedded systems is frequently an afterthought, as they rarely include clean, general-purpose interfaces for control or data exchange with a microprocessor, which complicates deployment in real-time, heterogeneous environments. 
Additionally, verification support is typically constrained to smoke tests, presenting challenges for researchers and engineers who need robust testing and validation pipelines for use in high-stakes scientific contexts. 
Our framework addresses these gaps by combining performance-efficient hardware with runtime programmability, a microprocessor-friendly interface for seamless system integration, and a comprehensive verification suite tailored to the demanding requirements of scientific edge computing.


\subsection{CGRA}

A Coarse-Grained Reconfigurable Array (CGRA) is a type of architecture that has seen recent resurgence in industry and academic designs. 
Commercial CGRAs include Samsung Reconfigurable Processor~\cite{samsungCGRA} and the Renesas STP Reconfigurable Processor~\cite{renesasSTP}.
CGRA-ME~\cite{chin2017cgra} is an exemplary academic framework that supports a subset of CGRA designs.

CGRAs were originally envisioned as coarse-grained FPGAs whose programmable elements work at the word-level instead of FPGA bit-level programming~\cite{cgraSurvey}. 
Early CGRAs were classified based on their integration into the processor.
Tightly coupled CGRAs integrate into a processor data path and are executed as a custom instruction, e.g., Chess~\cite{marshall1999reconfigurable}, MATRIX~\cite{mirsky1996matrix}, and DySer~\cite{govindaraju2012dyser}. 
Loosely coupled CGRAs act more as an accelerator that executes alongside the processor, executing in tandem with the processor and communicating via on-chip interconnect. 
Examples of loosely-coupled CGRAs include PipeRench~\cite{goldstein2000piperench}, MorphoSys~\cite{singh2000morphosys}, CHARM~\cite{cong2012charm}, and FPCA~\cite{cong2014fully}.
\codename can be characterized as a loosely coupled CGRA as it implements sub-tasks of the neural network alongside a CPU with a focus on enabling hardware/software partitioning.

Modern neural network accelerators often exhibit architectural similarities to CGRAs, particularly in their use of spatial dataflows, on-chip communication networks, and pipelined compute units. 
Like CGRAs, many NN accelerators rely on configurable interconnects and local memory structures to maximize data reuse and throughput. However, neural network accelerators are typically more narrowly optimized for specific computational patterns and may hardwire aspects of control or scheduling for performance. 
In contrast, CGRAs are generally more flexible—supporting a broader range of computational kernels and allowing dynamic reprogramming of data paths at runtime. 
With \codename, we aim to navigate the design space between these two classes: delivering an architecture that captures the efficiency and specialization of NN accelerators while retaining the programmability and reuse benefits of CGRAs.

\revision{
In the recent literature, OpenCGRA~\cite{opencgra} introduced a unified, open-source stack to model, test, simulate, and characterize CGRAs, with LLVM compiler passes and PyMTL-based Verilog generation, such that designers can quickly explore heterogeneous tiles and full EDA flows in one place. 
AURORA~\cite{aurora} then automated CGRA co-design: starting from a generic architecture template, it jointly explores composition of functional units, memory/interconnect, and loop optimizations using an architecture-aware simulated annealing search. 
VecPAC~\cite{VecPAC} proposes a hybrid CGRA supporting different datatypes with scalar tiles plus configurable vector tiles.
PICACHU~\cite{picachu} introduces a plug-in CGRA for LLM nonlinear ops (GELU/Softmax/LayerNorm, etc.) with domain-specialized PEs, precision-aware design, and seamless integration with a systolic array.
ML-CGRA~\cite{MLCGRA} is an MLIR-based end-to-end framework that keeps ML-level semantics and adds passes lower whole models onto CGRAs.
AHA introduces an agile CGRA-compiler co-design flow where three domain specific languages: PEak for PEs, Lake for memories, and Canal for interconnects, auto-generate both RTL and compiler collateral from a single source of truth, so the compiler updates automatically as hardware evolves.
While such CGRA frameworks primarily target architecture exploration and compiler lowering, \codename's use case is complementary and deployment-oriented. 
It delivers vendor-agnostic SystemVerilog RTL and a production C runtime verified in-the-loop via DPI-C, packaged as AXI IP with DMA that can be integrated with SoC platforms such as Zynq, NanoSoC and Ibex, enabling rapid, reproducible deployment in heterogeneous SoCs. 
}

\newcommand{\Chlsml}{\cite{hls4ml}}
\newcommand{\Cfinn}{\cite{umuroglu2017finn}}
\newcommand{\Cvitis}{\cite{vitisai}}
\newcommand{\Cvta}{\cite{vta}}
\newcommand{\Cdlp}{\cite{dlp}}
\newcommand{\Copenvino}{\cite{openvino}}
\newcommand{\Ccaffiene}{\cite{zhang2023caffeine}}
\newcommand{\Ccgrame}{\cite{chin2017cgra}}
\newcommand{\Cflexcnn}{\cite{flexcnn}}
\newcommand{\chk}{\checkmark}
\newcommand{\SIM}{$\sim$}
\renewcommand{\arraystretch}{1.2}

\begin{table*}
    \centering
    \caption{Comparing \codename with existing ML-to-FPGA/ASIC frameworks. 
    \revision{As a deployment-oriented framework, \codename's unique feature is its comprehensive verification approach, which tests the user's model and the production runtime on the generated hardware via DPI-C with randomized SoC-level congestion emulation. In addition, we provide a superset of features from other frameworks in one place for the scientific computing community}}
    \label{tab:frameworks}
    \resizebox{\textwidth}{!}{%
    \setlength{\tabcolsep}{10pt}
    \begin{tabular}{rccccc} 
    \toprule
                                   &  This work & \hlsml \Chlsml & VTA  \Cvta & FINN \Cfinn  & Vitis AI \Cvitis \\  \midrule
Open-source                        &  \chk      & \chk           & \chk       & \chk         &                  \\
Sub-8-bit quantization support     &  \chk      & \chk           & \chk       & \chk         &                  \\
Larger neural networks support     &  \chk      &                & \chk       &              &  \chk            \\
Xilinx FPGA support                &  \chk      & \chk           & \chk       & \chk         &  \chk            \\
Intel FPGA support                 &            & \SIM           &            &              &                  \\
ASIC Implementation                &  \chk      & \SIM           &            &              &                  \\
Basic verification                 &  \chk      & \chk           & \chk       & \chk         &  \chk            \\ 
Generate production-ready runtime  &  \chk      &                & \chk       &              &  \chk            \\ 
Verification with runtime          &  \chk      &                &            &              &                  \\ 
    \bottomrule
    \end{tabular}
    }
    
    \vspace{1em}
    
    \resizebox{\textwidth}{!}{
    \begin{tabular}{rccccc} 
    \toprule
                                   &  DLP \Cdlp & OpenVino \Copenvino & Caffeine \Ccaffiene & CGRA-ME \Ccgrame & FlexCNN \Cflexcnn\\ \midrule
Open-source                        &            & \chk                &                     &  \revision{\chk}             & \chk    \\
Sub-8-bit quantization support     &            &                     &                     &                  &         \\
Larger neural networks support     &  \chk      & \chk                &   \chk              &  \chk            & \chk    \\
Xilinx FPGA support                &  \chk      &                     &   \chk              &  \chk            & \chk    \\
Intel FPGA support                 &  \chk      & \chk                &                     &  \chk            &         \\
ASIC Implementation                &            &                     &                     &  \chk            &         \\
Basic verification                 &  \chk      & \chk                &                     &  \chk            & \chk    \\ 
Generate production-ready runtime  &            & \chk                &                     &                  &         \\ 
Verification with runtime          &            &                     &                     &                  &         \\ 
    \bottomrule
    \end{tabular}
    }
\end{table*}

\subsection{ML to FPGA/ASIC Frontend Frameworks}
\label{subsec:frameworks}

There are many end-to-end frameworks available for hardware implementation of neural networks~\cite{surveyOfTools}.  
DNNBuilder~\cite{zhang2018dnnbuilder}  and \finn~\cite{umuroglu2017finn} implement a given model as a pipeline of layers, similar to \hlsml.
DNNBuilder allows the users to customize two kinds of reuse factors: channel and kernel. 
\finn takes in an ONNX model, possibly exported from Brevitas~\cite{brevitas},  to generate Vivado HLS IP. 
This can be verified in simulation before implementing on AMD Xilinx FPGAs. 
FINN provides a PYNQ driver for prototyping.

AMD Xilinx Vitis AI~\cite{vitisai} is a closed-source library that implements a Deep Learning Processing Unit, an 8-bit micro-coded processor to process neural networks optimized through their stack on AMD Xilinx FPGAs. 
OpenVino is a similar stack for Intel FPGAs.
Apache TVM, an open-source framework for embedded AI, implements a Versatile Tensor Accelerator (VTA)~\cite{vta} as a GEMM processor using Xilinx HLS.
LeFlow~\cite{leflow} emits HLS code, which has similar advantages and disadvantages as \hlsml.

\revision{
Recent compiler stacks such as PyTorch Dynamo~\cite{torchdynamo} and OpenAI Triton~\cite{openaitriton} focus on automatic kernel fusion and code generation for GPUs/CPUs (with early explorations for other backends). 
Our contribution is orthogonal: \codename targets a CGRA-style spatial accelerator with unified dataflow and a firmware/runtime that binds those schedules to a memory-mapped, DMA-driven engine on FPGA/ASIC. In principle, TorchInductor/Triton-style schedules could be lowered to our ISA/dataflow with a translation layer; we leave such integration to future work.
}

While each of these frameworks targets different kinds of users, the limitations in their approaches make them incompatible with the requirements of the scientific computing community, as listed in Table~\ref{tab:frameworks}. 
The popularity of \hlsml and its features: arbitrary quantization, Xilinx and Intel FPGA backends, and support for limited verification, demonstrate the unique needs of the scientific edge community. 
The models used in scientific applications require quantization to arbitrary bit-widths, which is made possible by \qkeras, and not supported by Vitis AI, OpenVino, or CGRA-ME, among others. 
In addition, their workflow involves implementing their models on FPGAs and later moving to ASIC designs, which is not possible with Vitis AI, Apache VTA, and OpenVino since their backends are implemented in vendor-specific HLS.

\subsection{Motivation for \codename}


Modern scientific applications increasingly rely on neural networks to process data rapidly and efficiently at the edge, imposing stringent requirements on throughput, latency, and power efficiency. 
\hlsml is a popular framework among scientists—deployed in custom ASICs for L1 triggers and on FPGAs in L2 triggers at the LHC, among other uses—because it offers an accessible, dataflow-style, layer-by-layer implementation for small networks and a range of precisions. 
However, this approach scales poorly beyond a few layers: on-chip buffering and per-layer specialization grow quickly, making larger models difficult to fit on FPGAs and small ASICs. 
Reusable accelerators amortize resources across layers and are therefore more suitable for larger networks, yet most available designs provide limited, non–user-friendly front ends and often lack features needed by the scientific computing community, including sub–8-bit quantization, multiple backends (FPGA and ASIC), end-to-end verification, and runtime generation. 
Moreover, \hlsml and other HLS-based DNN frameworks are tied to vendor-specific toolchains, which constrains backend portability and demands substantial engineering to build complete systems with a host CPU, DMA subsystems, drivers, and production firmware.

As summarized in Fig.~3, \hlsml remains the community’s go-to for dataflow implementations over varying bitwidths and shallow-to-moderate depths, while \finn likewise follows a dataflow paradigm and excels at ultra–low-precision designs. Traditional AI accelerators, in contrast, target large models with $\geq$8-bit quantization. \codename is designed to fill the gap between these regimes by enabling models with sub–8-bit quantization that are too large to realize with \hlsml or \finn, while providing multiple backends, end-to-end verification, and automatic runtime/driver generation to streamline full-system integration.

\section{\codename Overview}
\label{sec:overview}

The needs of the scientific computing community for a framework that can implement high-performance and programmable accelerators using the familiar \hlsml-style interface, with the generation of cross-platform hardware and production-ready firmware, were identified in Section~\ref{sec:background}.
Often models exceed on-chip memory capacities, demanding efficient off-chip data handling and more flexible hardware configurations.
Consequently, users often avoid larger models because \hlsml cannot implement them.
To address these limitations, we introduce \codename, an open-source modular framework explicitly designed for the scientific edge computing community that enables reuse between layers, implements larger models and eases integration into larger systems. 
\codename leverages the strengths of \hlsml: ease of use, quantization-aware training, and rapid hardware accelerated deployment, while substantially extending capabilities in several key areas. 

Firstly, \codename supports large-scale neural network implementations by employing a parametrizable CGRA, enabling efficient spatial reuse of processing elements and optimized off-chip data movements.
\codename outputs SystemVerilog RTL eliminating the need for HLS tools required by other frameworks.
This RTL-centric approach streamlines FPGA implementation and facilitates seamless ASIC integration, thereby providing flexibility across different hardware targets without extensive code modifications. 
Furthermore, \codename includes an extensive verification framework integrated within its workflow, significantly reducing the time spent in debugging and validation processes, which are notoriously time-consuming in hardware design.
This section describes the user workflow of \codename and the Python frontend that makes our infrastructure possible.

\begin{figure}
    \centering
    \includegraphics[width=0.7\columnwidth]{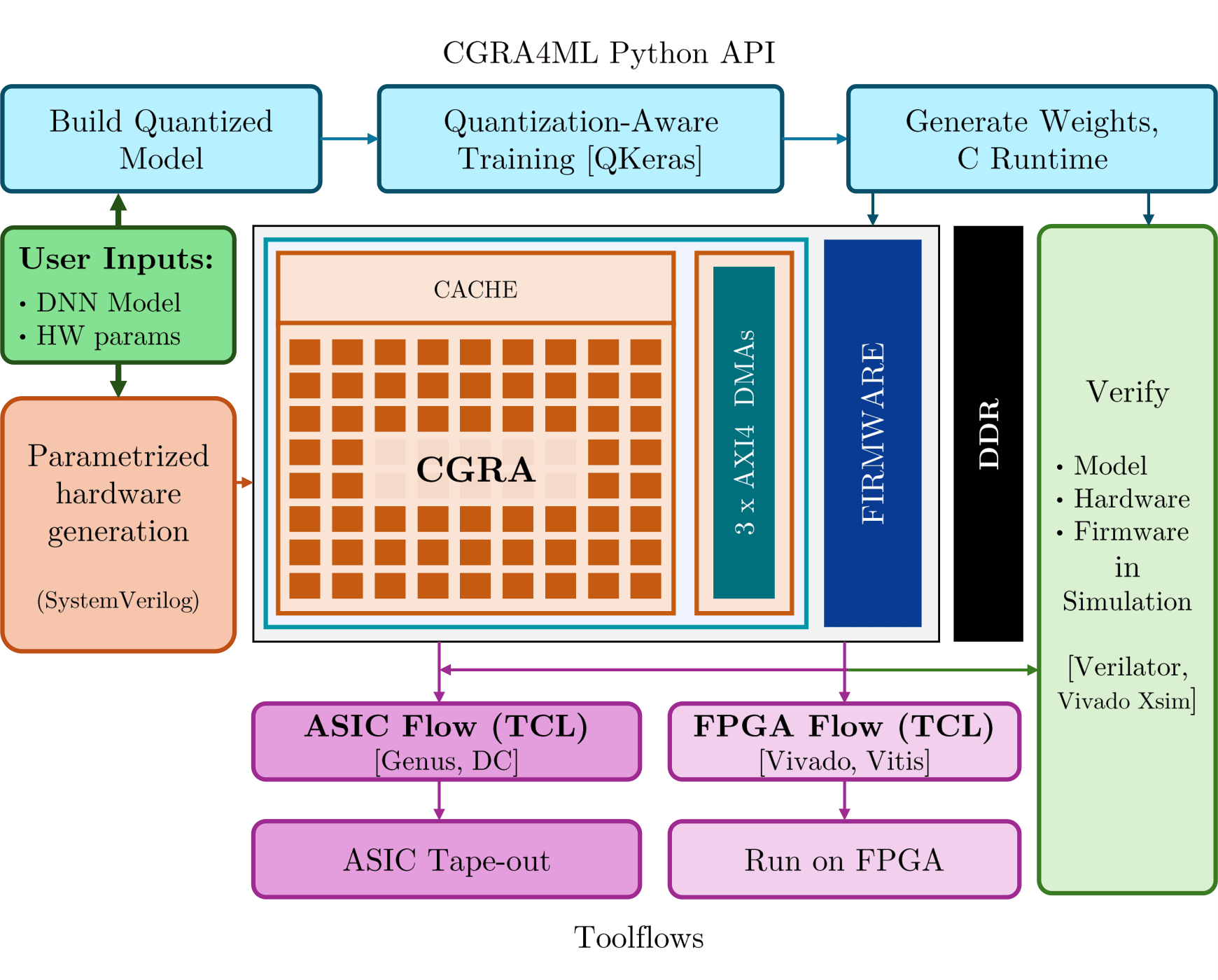}
    \Description{\codename workflow}
    \caption{\codename workflow as outlined in Sec. \ref{sec:overview}. 
    \label{fig:overview}
    Users first build quantized neural networks and train them in a quantization-aware manner using \qkeras ~\cite{qkeras} (Sec.~\ref{subsec:frontend}). Users then define a CGRA definition, which generates vendor-agnostic SystemVerilog RTL hardware specifications and TCL tool flows (Sec.~\ref{subsec:frontend}). The model can then be exported to generate weights and a C runtime firmware (Sec. \ref{sec:firmware}). The generated hardware IP is then verified comprehensively with the model and firmware (Sec. \ref{subsec:verf}) using our randomized, transactional SystemVerilog testbench suite with DPI-C extensions. Finally, the bitstream generated from the FPGA toolflow and the C firmware can be loaded into an FPGA to be tested in seconds (Sec. \ref{sec:fpga}). After such rapid prototyping, the same hardware design can be moved to ASIC (Sec. \ref{sec:asic}).}

\end{figure}


The overarching goal of \codename is to empower scientists and hardware designers by abstracting away intricate hardware optimization details that slow down development for non-expert users. 
\codename's user workflow is intuitive, streamlined, and flexible, supporting rapid development cycles and iterative experimentation. 
Users define neural network architectures via a Python API leveraging \qkeras for quantization-aware training, after which \codename automatically generates hardware specifications, runtime firmware, and robust verification suites. This end-to-end approach, shown in Fig.~\ref{fig:overview} enables rapid prototyping, flexible experimentation, and efficient deployment, ultimately allowing users to prioritize innovative applications over hardware intricacies.
It comprises the following steps:

\subsubsection{Model Definition and Training}
Users begin by defining neural network models using the Python-based \qkeras library, which supports quantization-aware training. 
This step ensures models are optimized for low-precision hardware implementations while maintaining accuracy.

\subsubsection{Accelerator Configuration}
After training, users specify static parameters for the CGRA architecture, including the number of processing elements (PEs), bitwidths, and memory hierarchies. 
Users can adjust these parameters to tune the design to within the resource constraints of the target hardware.

\subsubsection{RTL Generation}
Using the specified parameters, \codename automatically generates vendor-agnostic SystemVerilog RTL code. 
This eliminates dependency on vendor-specific tools and simplifies porting between FPGA and ASIC targets.

\subsubsection{Firmware Generation}
\codename produces optimized runtime firmware, facilitating seamless integration and operation of the generated hardware within an SoC environment. 
Users can quickly deploy and test their designs in real-world applications.

\subsubsection{Verification}
The generated hardware and firmware undergo comprehensive verification using \codename’s built-in randomized transactional testbench suite. 
This step ensures functional correctness and reliability prior to physical deployment.

\subsubsection{FPGA/ASIC Deployment}
Finally, users deploy their verified designs onto FPGA platforms for rapid prototyping and experimentation. 
Once validated, the same RTL design seamlessly transitions to ASIC flows.

Through this structured, \hlsml-inspired workflow, \codename enables users to rapidly move from model concept to physical deployment, streamlining innovation, and accelerating scientific discovery.

\section{\codename: Infrastructure}
\label{sec:infra}

To bridge the gap between high-level neural network development and hardware-accelerated edge deployment, \codename provides a modular, holistic infrastructure that spans model quantization, hardware generation, firmware synthesis, and system-level verification. 
Central to this flow is the concept of bundles, which are modular, deterministic units of execution that encapsulate groups of DNN layers, making them suitable for hardware acceleration. 
These bundles allow for clean partitioning of compute workloads between a coarse-grained reconfigurable array (CGRA) and a general-purpose CPU. This design philosophy supports scalability, reuse, and extensibility, enabling developers to experiment with diverse model structures while maintaining control over system-level performance and resource utilization. 
The backend pipeline, which comprises SystemVerilog CGRA RTL, portable runtime firmware, and automated toolchains for FPGA and ASIC flows, ensures that scientific edge applications can transition from simulation to physical deployment with minimal effort.

\subsection{Frontend Infrastructure: Model to Bundles}
\label{subsec:frontend}

\begin{figure}
    \centering
    \includegraphics[width=0.8\columnwidth]{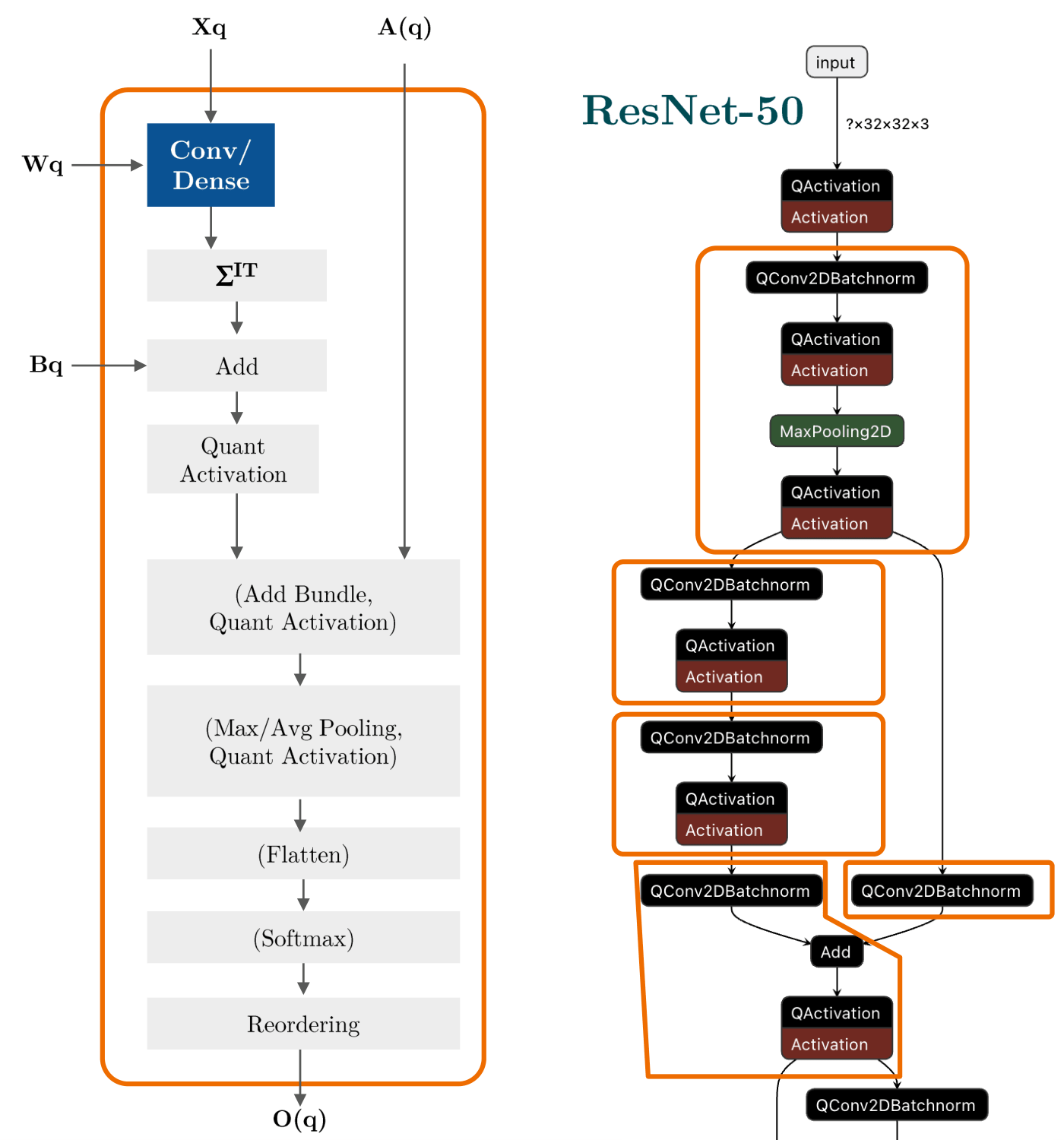}
    \caption{\textbf{Bundle:} A given DNN is decomposed into a list of bundles, where a bundle is a group of layers that can be deterministically executed by the system generated by \codename. The CGRA accelerates the simple but compute-heavy operations, while the CPU executes the complex but lightweight pixel-wise operations. This flexibility allows us to add new, complex operations easily.
    }
    \Description{A given DNN is decomposed into a list of bundles}
    \label{fig:bundle}
\vspace{-15pt}
\end{figure}

Modern neural networks have complex, seemingly arbitrarily structured graphs of computational layers.
Therefore, a framework should balance the flexibility it offers to the users to the optimized backend implementation of these networks in hardware.
This requires a modular and extensible intermediate representation (IR).
The frontend of \codename is responsible for translating a high-level neural network model into a hardware-ready specification. 
This process is centered around the concept of \emph{bundles}, which are modular, hardware-executable units that serve as the bridge between \qkeras models and \codename accelerator.

Users construct and train quantized models using \qkeras, a widely adopted library from Google that enables fine-grained control over quantization levels and supports straight-through estimators for backpropagation. 
Once trained, \codename parses the model’s computational graph and decomposes it into a list of bundles. 
Each bundle represents a deterministic sequence of layers that can be directly mapped onto the CGRA.
In the Python frontend, a bundle is implemented as a subclass of a \qkeras layer, enriched with metadata and configuration parameters required for hardware mapping. 
In the backend firmware, each bundle corresponds to a structured set of runtime instructions, including memory access patterns, kernel parameters, and execution schedules.

The \codename infrastructure is organized around the concept of bundles. 
They serve as modular, deterministic units of execution that explicitly define the boundary for hardware/software partitioning. 
Within each bundle, compute-intensive operations--specifically convolutions and dense layers--are offloaded to the CGRA to leverage the parallel multiply-accumulate (MAC) capabilities of its parameterized $R{\times}C$ processing element array. 
Conversely, lightweight or control-heavy tasks, such as pooling, normalization, and pixel-wise edge conditions, are allocated to the host CPU. 
This partitioning strategy is primarily motivated by the need to minimize hardware complexity; by offloading irregular control logic to the CPU, the CGRA remains optimized for high-density, regular computation. 
While the \codename frontend automatically parses computational graphs into these default partitions, the framework is intentionally user-definable; developers can group layers into custom bundles via the Python API to facilitate specific architectural patterns, such as residual skip connections, ensuring that new or non-standard operations can be deployed without modifying the underlying RTL.


Figure~\ref{fig:bundle} illustrates how a ResNet-50 model is decomposed into bundles, each encompassing layers such as Conv2D, activation, pooling, flattening, and skip connections. 
The Python frontend transforms the model, compares intermediate results against \qkeras outputs, and partitions execution between the CGRA and CPU. 
It selectively enables layers, and passes their configuration as runtime parameters inferred from the user defined model. 
The generated C firmware then orchestrates bundle execution at runtime in a way that ensures functional equivalence to the Python model. 

\codename also empowers users to define new bundles by extending the Python frontend and adding corresponding runtime definitions in C. 
These composite operations are decomposed during compilation into a static runtime, allowing users to define and deploy more complex architectures such as residual networks without modifying the core hardware design.
This architectural decision is made to keep the hardware complexity minimal, and to enable new operations such as activation functions when necessary. 
We characterize the CPU execution latency cost in Table~\ref{tab:cpuperf} and outline our future roadmap to move the writeback stage into hardware in Sec. \ref{sec:perfcpu}.

\subsection{Backend Infrastructure: CGRA RTL and Firmware Generation }

\begin{figure}
    \centering
    \includegraphics[width=0.7\columnwidth]{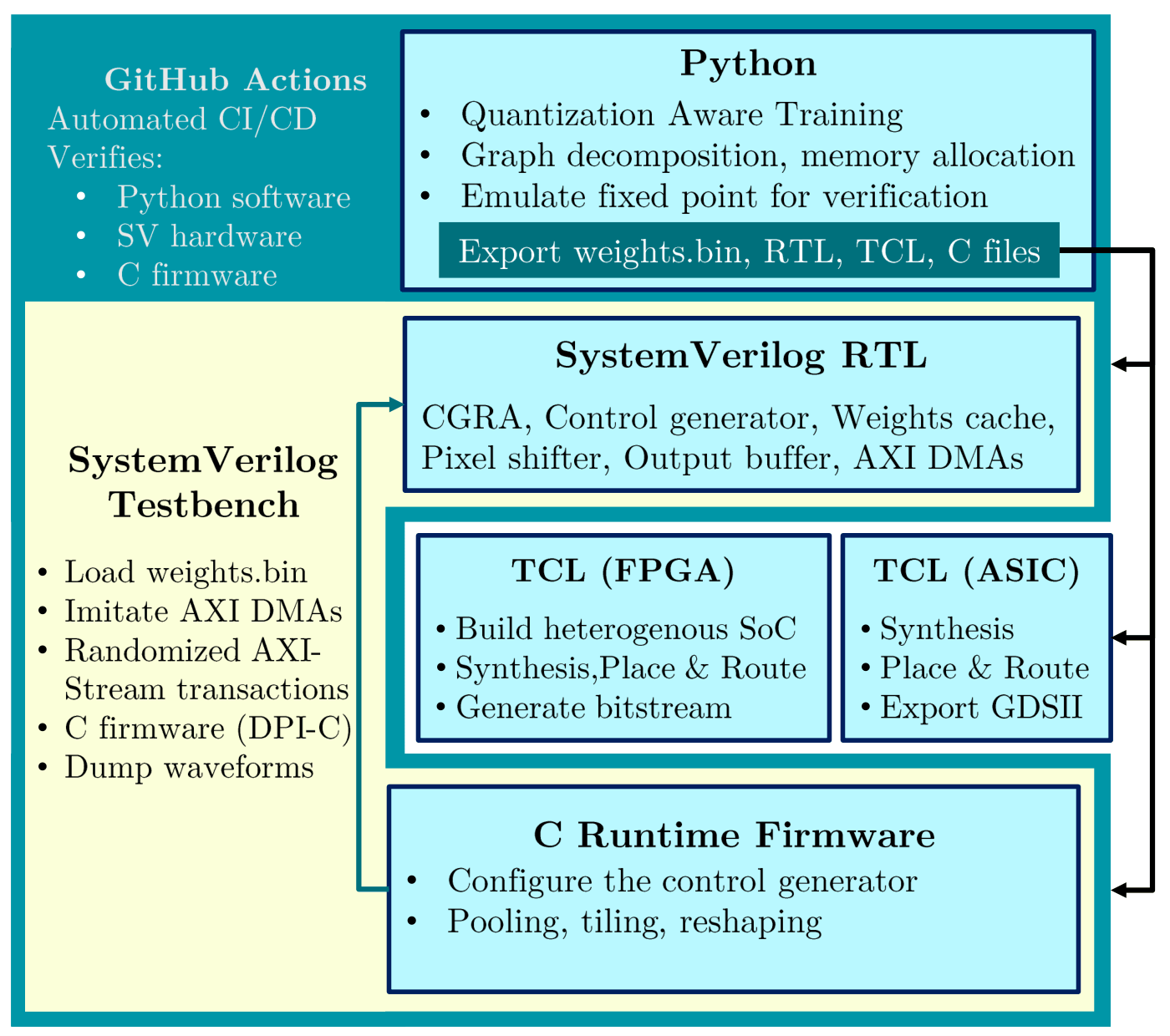}
    \caption{ Modular and extensible infrastructure of the \codename framework as described in Section \ref{sec:infra}. The Python API built around Google's \qkeras, extracts the fixed-point representations of intermediate tensors, generating the binary weights and model specification. The Python frontend also generates vendor-agnostic, synthesizable SystemVerilog RTL design of the CGRA per user specifications, optimized C runtime firmware and TCL toolflows for ASIC and FPGA implementations. Our comprehensive verification suite powered by randomized, transactional, SystemVerilog testbenches interfacing with the C firmware via DPI-C, then verifies the user's model, the generated CGRA, and runtime together as it would run on an embedded system, emulating cases like memory congestion. The CI/CD pipeline tests the entire framework on a set of models over different CGRA configurations}
    \Description{Modular and extensible infrastructure of the \codename framework as described}
    \label{fig:infra}
\end{figure}

The backend of \codename translates the high-level bundle specifications into hardware and runtime components, enabling deployment across FPGA and ASIC platforms. 
It encompasses three primary components: RTL generation, firmware synthesis, and toolchain integration for verification and implementation.
This subsection outlines these steps in the backend process, while the following subsections discuss each in detail.

\subsubsection{RTL Generation}
\codename produces synthesizable, vendor-agnostic SystemVerilog RTL for a parametrized CGRA engine. 
Users configure key static parameters, such as the number of processing elements, memory hierarchy depth, and data bitwidths, via the Python frontend. 
These parameters are embedded into the hardware description to generate an optimized datapath and control logic. 
The default datatype is fixed-point.
However, the users may integrate a floating-point MAC by specifying its latency of the multiplier and accumulator in our parameter list.
The resulting RTL supports standard AXI interfaces for seamless SoC integration.
After implementation, the design must be integrated into a broader SoC. 
\codename generates a high-performance, AXI-compliant IP powered by open-source, design tool-agnostic AXI DMAs. 
These modules simplify integration into custom SoCs or evaluation platforms. 
The firmware and hardware together ensure efficient data transfer and workload sharing between the CGRA and the host processor.

\subsubsection{Firmware Generation}
\label{sec:firmware}
Alongside RTL, \codename generates modular C firmware that manages CGRA execution and coordinates software-side tensor operations. 
This firmware is structured in three layers: (1) a model-specific configuration header exported from Python, (2) a portable runtime layer for buffer management and synchronization, and (3) architecture-specific routines for platform integration. 
The firmware is designed as a simple header-only library with a thin Hardware Abstraction Layer (HAL), allowing users to quickly get the design working and extend it to include co-processors and memory interfaces like Ethernet and PCIe.
\revision{This modularity allows the firmware to target diverse embedded processors including ARM, RISC-V, and x86, and to be reused across different CGRA configurations. 
The runtime targets a stable hardware contract via a custom bank of 32-bit registers for control and status, with DMA descriptors.
No changes to the unified dataflow or runtime firmware are required when targeting different architectures and FPGA/ASIC backends.
Only memory macro selection and interconnect adapters vary by process and SoC fabric.
We have validated the portability of our firmware on ZYNQ Programmable SoCs with FPGAs and ARM Cortex processors, on Ibex SoC with a RISC-V processor, and on x86 CPUs for verification. 
}

\subsubsection{Toolchain Integration}
\codename includes fully automated toolchains for FPGA and ASIC flows. 
For FPGAs, it generates TCL scripts that configure IPs, integrate the CGRA into SoC shells (e.g., Zynq), and synthesize/place/route the design. 
For ASICs, it produces scripts for Cadence Genus and Innovus, or Synopsys DC and ICC, enabling synthesis, physical design, and GDSII export. 
Users can customize these flows by linking to their PDKs, memory compilers, and setting specific constraints. 
Reports generated through these tools provide power, performance, and area (PPA) estimates. 
A proof-of-concept SoC using \codename IP integrated into the ARM-based NanoSoC platform demonstrates the viability of this flow.

\subsubsection{System Verification}
A central component of the backend is the DPI-C-enabled verification suite. 
As shown in Fig.~\ref{fig:firebridge}, testbenches written in SystemVerilog interface with the C firmware to simulate the complete stack: from quantized model outputs to hardware response. 
Randomized AXI stimuli stress-test corner cases such as memory congestion or timing edge conditions. 
This approach reduces the verification burden and ensures fidelity between the model and its physical implementation.

Together, these backend systems allow \codename to bridge the gap between machine learning model development and production-grade hardware deployment, offering flexibility, correctness, and performance portability across platforms.

\subsection{Parameterized CGRA Engine Architecture}
\label{subsec:parametrize}

\begin{figure}
    \centering
    \includegraphics[width=0.6\columnwidth]{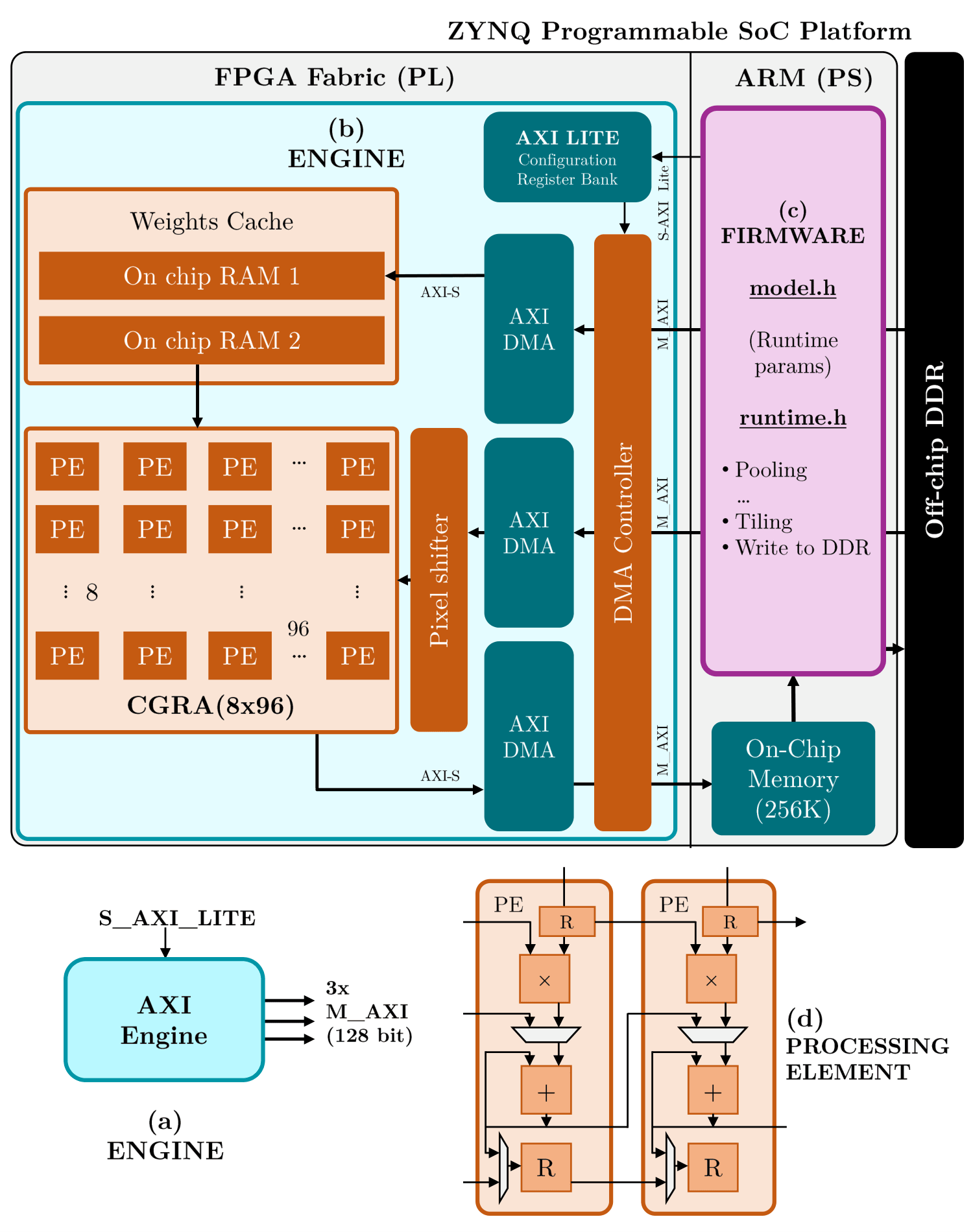}
    \caption{\codename architecture generated for AMD Xilinx Zynq SoC: Our CGRA-based engine \textbf{(a)} features one AXI-Lite port for configuration and three 128-bit AXI Manager ports for data movement. 
    The CGRA engine \textbf{(b)} performs compute-heavy tasks. 
    The engine has parameterizable bit widths, weights cache, and processing elements (PE) to fit the requirements of different models and meet hardware resource and performance constraints. 
    \codename emits software for the ARM processor \textbf{(c)} to perform pixel-wise operations and handle the edge cases. 
    }
    \Description{\codename architecture generated for AMD Xilinx Zynq SoC}
    \label{fig:sys}
\end{figure}

\codename generates a system of hardware components, as described in Fig.~\ref{fig:sys}, that work together with a host processor to process the user defined neural network model.
The components are designed to have AXI interfaces for modular hardware design.
The system consists of the following subsystems, each with specified parameters that the user sets using the Python API:

\begin{itemize}
    \item CGRA - A Coarse Grained Reconfigurable Array (CGRA) with AXI-stream interfaces. 
        \begin{itemize}
            \item Number of Rows \& Columns of processing elements (PEs)
            \item Bitwidth of inputs, kernel, outputs, and bias
        \end{itemize}
    \item Weights cache - an AXI stream module with ping-pong buffers of on-chip SRAM for maximizing data reuse, counters and controllers.
        \begin{itemize}
            \item Depth of weights cache
            \item Maximum batch, kernel sizes, in-channels, height, and width
        \end{itemize}
    \item Three Direct Memory Access (DMA) modules that convert between AXI stream and full AXI4.
        \begin{itemize}
            \item Bitwidth of AXI interfaces (up to 128-bits)
        \end{itemize}
    \item Verification IPs and wrappers for full system Verification
        \begin{itemize}
            \item Valid and Ready probabilities for randomization
        \end{itemize}
\end{itemize}
In addition to these tunable parameters, clock frequency and I/O delays of the full system are also parametrizable.

\codename implements large and diverse neural networks by creating an efficient yet programmable computational array that can easily be shared across multiple neural network layers. 
The array's computation and data movement are design-time parameterizable and run-time programmable to accommodate a variety of DNN layer types. 
The CGRA is designed to perform various types of neural network layers with optimal performance.
This subsection discusses the static, design-time parametrization of the CGRA, while the next subsection discusses runtime configuration.

The static hardware specification defines the CGRA architecture, which can be reprogrammed at runtime within these constraints to execute many different models. 
These attributes are parameterized into the SystemVerilog hardware to generate different CGRA engines quickly. 
The parameters are made static before synthesis based on the user-defined attributes. 
\codename supports fixed-point datatypes, since quantization is critical to meet high throughput and low latency requirements in scientific applications. 
Their bitwidths are customizable across the inputs, outputs, kernel, and bias.
The PE datapaths are automatically adapted to match the requirements of these data types. 

The rows and columns dictate the number of PEs in the CGRA, which define the hardware resource usage and the peak performance. 
Fig.~\ref{fig:sys} (d) shows the internal architecture of PEs, which include a multiplier, accumulator, two registers, and two multiplexers. 
Simplicity is favored over complexity to increase computational density and efficiency. 
CGRAs designed with a higher degree of flexibility in data movement and operation types suffer from the complexity required to program them and often utilize their resources less effectively.
Therefore, our lightweight CGRA design aims to offer minimal programming complexity while supporting high utilization required for scientific edge computing. 
The PE columns can group themselves and process convolution kernels, sharing data to maximize data reuse.

The depth of the weights cache controls the weight reuse factor. 
The weights cache consists of two on-chip RAMs with parameterized depths in a ping-pong configuration to allow full throughput.
The pixel shifter maximizes data reuse across vertical convolution while maintaining modularity.
Their dimensions provide a trade-off between performance and available hardware resources.
While any model can be run on any specification, the hardware parameters can be optimized for specific models as described in Subsection~\ref{sec:perf}. 
The maximum limits of layer dimensions determine the register sizes within the hardware (see Table~\ref{tab:runtime_params}). 

As shown in Fig.~\ref{fig:sys}, our hardware modules feature open AXI-Full, AXI-Stream, and AXI-Lite interfaces to ensure compatibility with a wide variety of SoCs and other hardware IPs. 
We utilize open-source AXI and AXI-Stream cores~\cite{alexaxi, alexaxistream} in our SoC generation, modified to support ASIC flow in addition to FPGA.
The AXI interfaces are parameterized to easily match the SoC interfaces. 
All these modules are wrapped into a single IP that can be easily instantiated and connected to any system with AXI ports.

\subsection{Runtime Dynamic Reconfiguration and Dataflow}
\label{subsec:reconfig}

\begin{table}
\centering
\begin{threeparttable}
\caption{Runtime parameters of the unified dataflow }
\begin{tabular}{l|l|l|l|l}
  \toprule
  \multirow{2}{*}  & \multicolumn{2}{c|}{Description}  & \multicolumn{2}{c}{Array Partition}   \\
         & Conv           & Matmul       & Slice (S)                                          & Iterations (T)\\
  \midrule
  $K_H$  & Kernel Height  & 1            &                                                    &               \\
  $K_W$  & Kernel Width   & 1            &                                                    &               \\
  $N$    & Conv Batch     & 1            &                                                    &               \\
  $W$    & In Im. Width   & 1            &                                                    &               \\
  $H$    & In Im. Height  & I/O Rows     & $H_S{=}R$\tnote{\textdaggerdbl}                    & $H_T{=}\lceil H/H_S \rceil$ \\
  $I$    & In Channels    & In Cols      & $I_S{=}\lceil D_W/K_H\rceil$\tnote{\textdagger}    & $I_T{=}\lceil I/I_S \rceil$ \\
  $O$    & Out Channels   & Out Cols     & $O_S{=}\lfloor C/K_W\rfloor$\tnote{\textdaggerdbl} & $O_T{=}\lceil O/O_S \rceil$
\label{tab:runtime_params}
\end{tabular}
\begin{tablenotes}
    \item[\textdagger]    $D_W{=}$ Depth of SRAM in weights rotator
    \item[\textdaggerdbl] $R,C{=}$ Rows \& columns of PE array
\end{tablenotes}
\end{threeparttable}

\centering
\includegraphics[width=0.6\columnwidth]{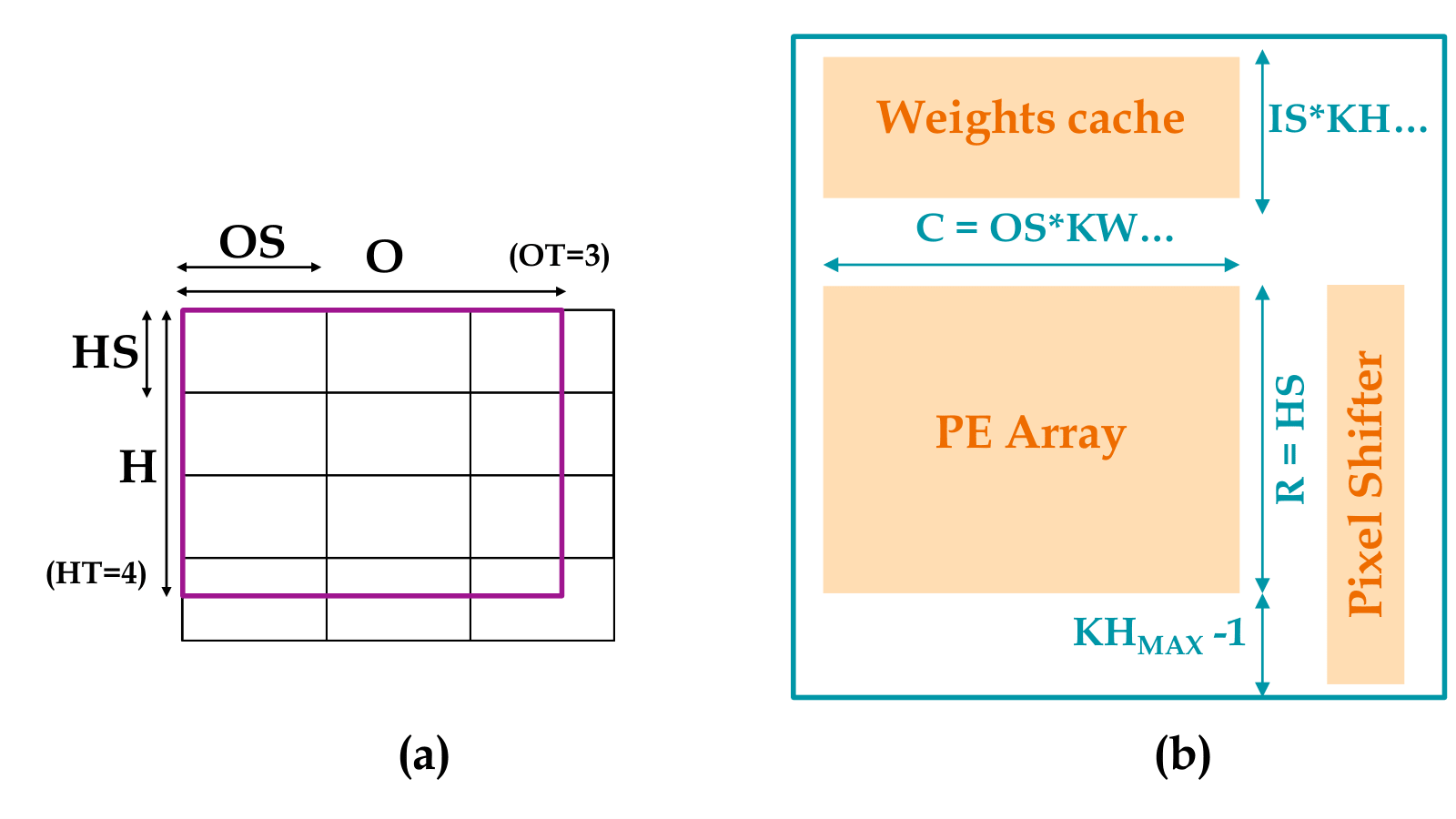}
\vspace{-10pt}
{
  \makeatletter
  \def\@captype{figure}
  \makeatother
  \caption{\textbf{Tiling:} An output matrix with a shape $[H,O]$ is generated as \underline{S}lices of shape $[H_S, O_S]$ (a), in $HT{\times} OT$ i\underline{T}erations by a CGRA. The compile-time hardware parameters: depth of weights cache, rows \& columns of CGRA ($R{\times}C$) affect the runtime parameters as shown in (b).}
  \Description{Matrix tiling on a Coarse-Grained Reconfigurable Architecture (CGRA). (a) illustrates an output matrix of shape $[H, O]$ being partitioned into smaller slices of shape $[H_S, O_S]$. These slices are processed over multiple iterations ($H_T \times O_T$). (b) shows how these parameters map to the static parameters of CGRA.}
  \label{fig:slice}
}

\end{table}

\begin{figure}
    \centering
    \includegraphics[width=\columnwidth]{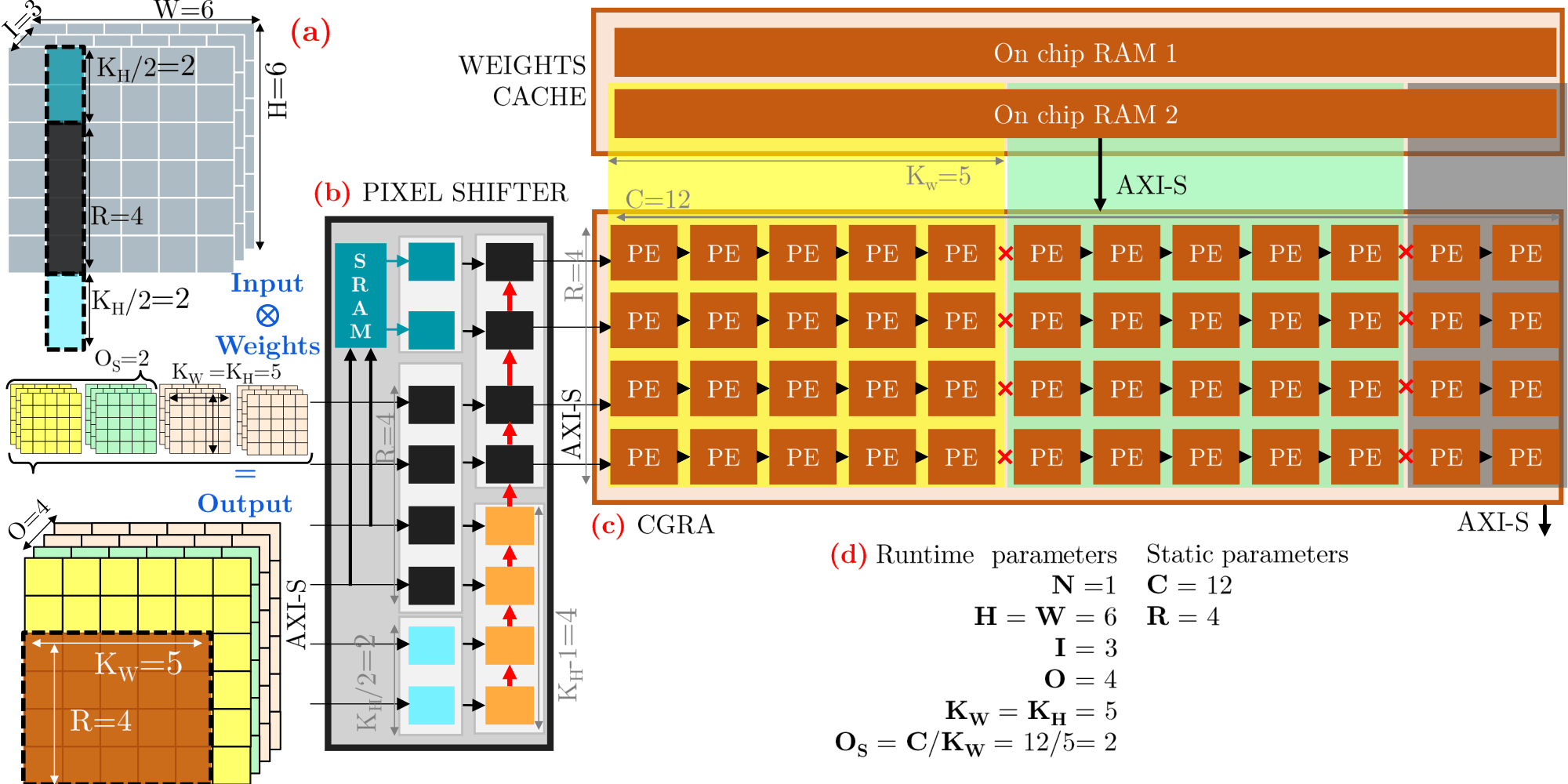}
    \Description{Convolution through Unified Dataflow}
    \caption{\textbf{Convolution through Unified Dataflow:}
    \revision{
  \textbf{(a)} An input tensor of shape $[N{=}1,H{=}6,W{=}6,I{=}3]$ is convolved with weights tensor of shape $[K_H{=}5,K_W{=}5,I{=}3,O{=}4]$ to generate an output tensor of $[N{=}1,H{=}6,W{=}6,O{=}4]$.
  The runtime parameters \textbf{(d)} are determined by the dimensions of the workload, while the dimensions of the hardware determine the static parameters.
  In this example configuration, the $R,C{=}4,12$ CGRA dimensions \textbf{(c)}  map to the slice of the output tensor as shown in \textbf{(a)}.
  The 12 columns of PEs dynamically group into 2 groups of 5 to process the horizontal convolutions ($K_W{=}5$).
  The two groups process two output channels ($O_S{=}2$) of the output tensor.
  The pixel shifter \textbf{(b)} receives 6 words of the input tensor: $R=4$ rows that correspond to the mapping in the output tensor, and $K_H/2{=}2$ rows below them.
  The last $K_H/2{=}2$ rows among the $R$ middle rows are stored in an on-chip SRAM to be used as the top rows when processing the next $H_S{=}R$ height slice. 
  The top $K_H/2{=}2$ rows needed for the current $K_H{=5}$ vertical convolution come from this RAM. 
  These $R{+}K_H-1{=}8$ words are loaded into a shift register bank and shifted 5 times.
  The $R$ rows of the CGRA receive the top $R$ rows of the shift register and thus obtain the vertical neighborhood needed for the convolution via shifting.
  After this shifting, the next input channel of the input image is loaded.
  After all input channels are loaded and shifted, the PEs of the CGRA (within a group) pass their accumulator values to the PE on their right to be accumulated there.
  The five PEs in a group (two groups are colored yellow and blue) perform the horizontal convolution this way.
  Since each group corresponds to an output channel, two such channels are processed in parallel.
  }
  }
    \label{fig:convdf}
\end{figure}

While the hardware architecture is defined by static design-time parameters, \codename provides further flexibility through dynamic runtime reconfiguration. 
This adaptability is driven by the runtime parameters detailed in Table~\ref{tab:runtime_params}, which allow the system to optimize its execution for various model architectures without hardware modification. 
These parameters specifically govern the Unified Dataflow and its associated tiling patterns, which are architected to maximize weight reuse and minimize costly off-chip data movement. 
By regulating the predictable movement of weights, inputs, and outputs during convolutions and dense layers, the framework ensures high hardware utilization. 
This dataflow is fully automated and managed via decentralized control logic distributed across the processing elements (PEs). 
As configuration bits flow through the array alongside the data, individual PEs can autonomously reconfigure their internal data paths at every clock cycle to process layers efficiently.


The Python API generates the configuration bits for a layer and passes them to the C firmware.
During the setup phase, the firmware writes attributes to the AXI-Lite configuration registers.
During the operation, the DMA controller reads the configuration bytes and passes them along with data through the AXI DMAs.
The Weights cache and Pixel Shifter receive these configuration bits, process them into fewer bits, and pass them along with the AXI-Streams as TUSER bits to the CGRA.
The multiplexers inside the PEs respond to these control bits and reroute the data dynamically at every clock cycle.
Since the configuration moves with the data, the CGRA PEs reconfigure themselves on the fly, grouping and sharing partial data to process a layer efficiently, without waiting to clear the pipelines.
The runtime parameters and the precise description of our default dataflow through the CGRA are presented in Table~\ref{tab:runtime_params} and Algorithm~\ref{alg:dataflow}, respectively.

Next, we describe the two specific cases of dataflow: Matrix Multiplication and Convolution.
The flexibility and modularity of the \codename framework enables additional use cases with minimal modification.
Additional layers and activation functions are added by modifying the C firmware.


\begin{algorithm*}
  \caption{Unified Dataflow}
  \label{alg:dataflow}
  \nonl\begin{minipage}[t]{0.97\linewidth}\footnotesize
  \revision{
  \textbf{Notation legend.}
  $(\cdot)_S$ and $(\cdot)_T$ denote \emph{slice} and \emph{iteration} partitioning, respectively.
  We map the $R{\times}C$ PE array to output height and out-channels via
  $H_S{=}R$, $H_T{=}\lceil H/H_S\rceil$, $O_S{=}\lfloor C/K_W\rfloor$, $O_T{=}\lceil O/O_S\rceil$.
  Input channels are processed with parallel factor $I_S$ and iterations $I_T{=}\lceil I/I_S\rceil$; in our design
  $I_S$ is chosen so $K_H\!\cdot\!I_S \le D_W$ (weights-SRAM depth), which gives 
  $I_T{=}\lceil D_W/K_H\rceil$ used below.
  We reorder tensors to expose reuse:
  $X\!\to\!X_T$ (vertical locality via the pixel shifter),
  $K\!\to\!K_T$ (row-major streaming into the weights cache).
  Braces labeled \emph{data beats} indicate DMA stream order; those labeled \emph{parallel words} indicate
  words consumed/produced per cycle by the array.
  }
  \end{minipage}
  \par\smallskip
  \nonl
  \begin{flalign*}
    X   &: [N,H,W,I] & \text{input}\\
    X_S &: [N,(H_T,H_S),W,(I_T,I_S)] \quad \text{where} \quad  H_T{=}\lceil H/R \rceil,H_S{=}R,I_T{=}\lceil D_W/K_H \rceil, I_S{=} \lfloor I/I_T\rfloor   & \text{slicing} \\
    X_T &: \underbrace{[I_T,N,H_T,W,I_S,}_\text{data beats} \underbrace{H_S{+}K_H{-}1]}_\text{parallel words} & \text{reordering} &
  \end{flalign*}
  \begin{flalign*}
    K   &: [K_H,K_W,I,O] & \text{weights}\\
    K_S &: [K_H,K_W,(I_T,I_S),(O_T,O_S)] \quad \text{where} \quad  O_S{=} \lfloor C/K_W \rfloor, O_T{=} \lceil O/O_S \rceil & \text{slicing} \\
    K_T &: \underbrace{[I_T,O_T,}_\text{(AXI-S Packets)} \underbrace{I_S,K_H,}_\text{(W-SRAM rows)} \underbrace{O_S,K_W]}_\text{(parallel words)} & \text{reordering} &
  \end{flalign*}
  \begin{flalign*}
    Y   &: [N,H,W,O] & \text{output}\\
    Y_S &: [N,(H_T,H_S),W,(O_T,O_S)]   & \text{slicing} &\\
    Y_T &: \underbrace{[I_T,O_T,N,H_T,W,}_\text{data beats} \underbrace{O_S,H_S]}_\text{parallel words} & \text{reordering} &
  \end{flalign*}
  \begin{flalign*}
    A^\prime &: [C,R] & \text{Accumulators of PE array}\\
    A        &: [O_S,K_W,R] & \text{dynamic regrouping}
  \end{flalign*}
  \hrulefill
  \SetKwFor{ParFor}{parallel for}{do}{end}
  \SetKwBlock{DoParallel}{do in parallel}{end}
  \SetInd{0.5em}{0.5em}
  
  $A[:,:,:] \gets 0$
  \BlankLine
  \nonl\For{$i_t < I_T (=I/I_S)$}{ 
    \nonl\For{$o_t < O_T (=O/O_S)$}{
      \nonl\For{$n < N$}{      
        \nonl\For{$h_t < H_T (=H/H_S)$}{    
\SetAlgoVlined
        \nonl\For{$w < W$}{
\SetAlgoNoLine
          \nonl\For{$i_s < I_S$}{ 
            \nonl\For{$k_h < K_H$}{
              \nonl\ParFor{$o_s < O_S (=C/K_W)$}{ 
                \nonl\ParFor{$k_w < K_W$}{        
                  \nonl\ParFor{$h_s < H_S (=R)$}{ 
                  \nonl $x \gets X_T [i_t, n, h_t, w, i_n, h_s{+}k_h]$ \\
                  \nonl $k \gets K_T [i_t,o_t,i_s,k_h,o_s,k_w]$ \\
                  \nonl $A[o_s,k_w,h_s] \mathrel{{+}{=}} k*x$ \\
}}}}}
\BlankLine
\nonl $Y[i_t,o_t,b,h_t,w,:,:] \gets A[:,{-}1,:] $ \\
\nonl \ParFor{$k_w < K_W$}{ 
        \nonl $A[:,k_w,:] \gets  (k_w \equiv 0)$ ? $0 :  A[:,k_w{-}1,:] $ 
}}}}}}
\end{algorithm*}

\subsubsection{Matrix Multiplication} 

Figure~\ref{fig:slice} shows the tiling (array partitioning), which is primarily output-stationary dataflow. 
When multiplying two matrices of sizes $[H,I]$ and $[I,O]$ to obtain an output matrix of size $[H,O]$, the output matrix is divided into slices of size $[H_S,O_S]$ to be processed in $H_T{\times}O_T$ iterations, such that $H_T{=}\lceil H/H_S \rceil$ and $O_T{=}\lceil O/O_S \rceil$. 
The rows and columns of the CGRA correspond to the size of a tile that can be processed at a time: $R{\times}C{=}H_S{\times}O_S$. 
When processing dense layers, the image batch is presented along $H$ to reuse weights across the batch.

\subsubsection{Convolution}

The unified dataflow is designed to maximize convolution data reuse. 
Fig. \ref{fig:convdf} describes a convolution workload through the unified dataflow with an example workload.
$R$ rows of processing elements compute $R{=}H_S$ Height slices. 
Algorithm \ref{alg:dataflow} describes the preparation of the input, weights and output tensors to be processed by the CGRA.
\revision{
The reuse-aware bandwidth model in Eqs.~\ref{eq:clocks}-\ref{eq:out} of Section \ref{sec:perf} quantitatively matches the schedule above and guides $R,C,I_S$ choices explaining the utilization/data-movement trends in Fig.~\ref{fig:perfres}. 
}

\revision{
An input tensor $X$ of shape $[N,H,W,I]$ is sliced according to the static parameters and reshaped into a tensor $X_T$ of shape $[I_T,N,H_T,W,I_S,R{+}K_H{-}1]$.
Since the output side has much lower bandwidth, slicing and reshaping the input for the next layer happens at the output side of the current layer, and the words are written to the respective locations in memory to form the tensor $X_T$. 
When processing the next layer, the DMAs read $X_T$ in row-major order (as written).
The pixel shifter takes $R{+}K_H{-}1$ values from the AXI DMA and shifts them, reducing the required input bandwidth by $K_H\times$ by exploiting data locality. 
}

\revision{
The weights cache has two buffers that work in ping-pong configuration, each $C$ words wide and $D_W$ rows deep.
The weights tensor $K$ of shape $[K_H,K_W,I,O]$ is sliced and reshaped into a tensor $K_T$ of shape $[I_T,O_T,I_S,K_H,O_S,K_W]$.
This slicing and reshaping is done offline, by the Python frontend, which then generates the weights binary that is loaded with the firmware, and read by the DMA in row-major order (as is).
The weights stored in the weights cache are rotated and reused $O_T N H_T W$ times, maximizing the weights reuse and reducing the required bandwidth proportionally.
}

\revision{
$C$ columns of PEs re-group dynamically into $O_S{=}\lfloor C/K_W \rfloor$ groups to process one Out-Channel slice $O_S$, while each such group of $K_W$ columns process the horizontal convolution, exploiting its data locality, and reducing input bandwidth by the same amount. 
$H_S {\times} O_S$ output pixels are computed in parallel by the $R{\times}C$ array of processing elements.
}

Taken together, these components form a robust and extensible foundation for deploying quantized neural networks in embedded scientific systems. 
By abstracting hardware complexity through modular design and automation, \codename allows scientists and developers to focus on model innovation without being encumbered by low-level engineering overhead. 
The ability to generate synthesizable RTL, runtime firmware, and verified execution flows ensures consistency across simulation, emulation, and deployment platforms. 
Moreover, the framework’s flexibility in supporting different processor architectures, SoC shells, and toolchains makes it well-suited to the rapidly evolving demands of scientific edge computing. 
The seamless integration of verification infrastructure, programmable CGRA engines, and composable bundles underscores the unique value of the framework as a research and deployment tool for ML-powered embedded systems.

\section{\codename: Verification and Deployment}

Robust verification and deployability are crucial for closing the loop between neural network design and reliable hardware execution, particularly in scientific edge applications, where correctness, reproducibility, and system integration are paramount. 
While many accelerator frameworks focus on performance or programmability, \codename places equal emphasis on comprehensive verification and streamlined hardware deployment. 
This section describes the verification methodology that ensures functional consistency across model, firmware, and RTL, followed by the deployment toolflows for FPGA and ASIC targets. 
Our infrastructure supports agile development and testing through simulation, CI/CD automation, and real-world deployment on programmable SoCs and standard ASIC flows, giving researchers and developers confidence that their designs will perform reliably across the entire development stack.

\subsection{Comprehensive Verification Approach}

\begin{figure}
    \centering
    \includegraphics[width=0.7\linewidth]{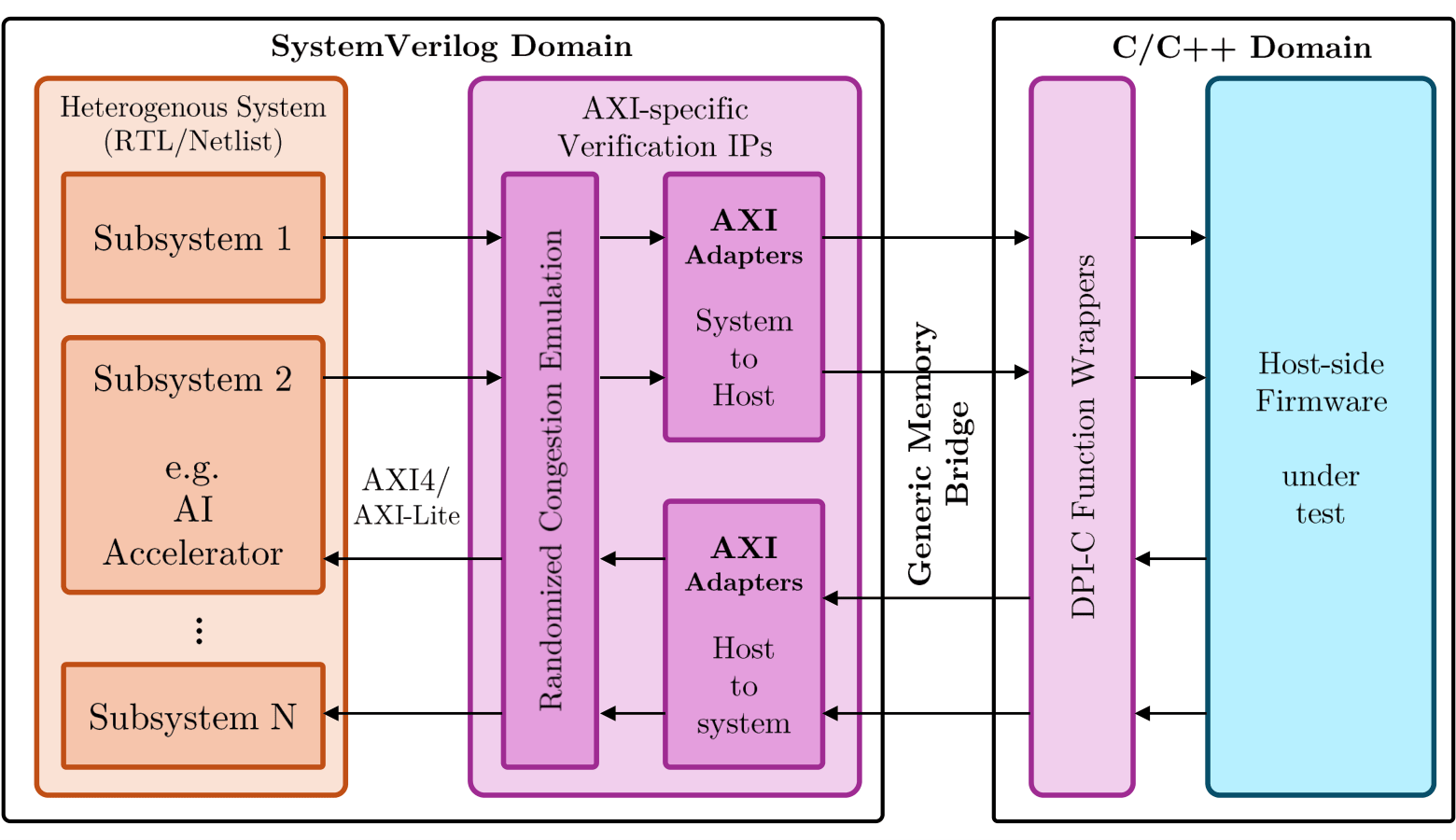}
    \Description{Verification Infrastructure developed for \codename}
    \caption{
    Verification Infrastructure developed for \codename. 
    The subsystems of CGRA, cache, DMAs and DMA controller are described in SystemVerilog, and during verification are connected to the production-ready C/C++ firmware through AXI-specific VIPs and custom DPI-C wrappers. 
    This way, the same firmware that runs on the FPGA also runs in simulation, verifying the entire system holistically.
    }
    \label{fig:firebridge}
\end{figure}

\label{subsec:verf}
Verification is an often overlooked aspect of research designs, which results in significant friction when transitioning from FPGA to ASIC flows.
When users implement sophisticated models and developers add new features, it is important to ensure they will work as intended on the final embedded system before investing the effort to implement them.
Our verification infrastructure has been tested on Cadence Xcelium, AMD-Xilinx Vivado Xsim, and Verilator (open source).

\codename provides a verification suite that is integrated across hardware, software, and firmware, \revision{targeting 100\% bit parity with a CPU/GPU golden run at the layer, bundle, and model outputs, using the exact numeric format used by the hardware.}
Our randomized, transactional testbench suite is built in SystemVerilog and interfaces with the IP's AXI-Manager ports.
The valid and ready probabilities define the randomness in toggling the \emph{s\_valid} and \emph{m\_ready} handshake signals of the AXI interfaces. 
Setting the probabilities to 1 gives the fastest simulation, where data is transferred every cycle. 
This is helpful as a smoke test when a user intends to modify the hardware and debug their modification. 
Setting it to lower values, e.g., 0.1 and 0.01, stress tests the hardware using randomized transactions, exposing hardware bugs that smoke tests cannot reveal.
The SystemVerilog Direct Programming Interface for C (DPI-C) is utilized in a novel manner to interface runtime firmware functions with the SV testbench suite to achieve holistic verification.

From the user's perspective, this verification process is streamlined as follows.
First, a randomized tensor is passed through the user's deep neural network model to capture intermediate and final output tensors in the software frontend. 
These are sliced, reshaped, and tiled to produce the expected inputs and outputs from the dataflow.
During simulation, the data memory resides on the C side and is accessed from the SV side through custom functions.
When the simulation begins, a function from the C runtime is called through DPI-C to load the model-specific configuration into the AXI-Lite register bank.
As the simulation progresses, the AXI interfaces of the SV testbench read the data memory from the C side and feed the tensors into the RTL design.
The outputs from the design are similarly stored into the memory.
This process is randomized, such that the valid \& ready signals of the address, data and response channels of the AXI interfaces are randomly toggled at the probability set by the user.
This randomization emulates the behavior of a congested bus interface, stress-testing the system under realistic conditions and revealing bugs that would otherwise be unidentifiable.
The intermediate and final results from the simulation are dumped into files, which the Python API then tests against the expected results.

Our comprehensive verification suite ensures that the model built and trained by the user works as it should on the RTL design and the firmware of the entire embedded system, before even starting the implementation process. 
\revision{All simulations are checked against a CPU/GPU golden reference with identical datatype to the RTL (same data type, per-tensor scale/zero-point, padding/stride/dilation, and the RTL’s rounding and saturation rules). 
The same checks can be executed on the FPGA build by streaming the identical tensors through the AXI DMA and comparing the captured outputs to the golden results, enabling end-to-end parity checks beyond simulation.}
After implementation, the same suite can be used on the gate level, time annotated netlist.
This allows the developer and the user to add functionality to hardware and the runtime firmware and test it on their desktop before testing it on an FPGA, allowing agile development cycles.
A set of GitHub actions running in the cloud verify the entire infrastructure using Verilator for multiple random examples after every update, enabling continuous integration and development (CI/CD), as shown in Fig.~\ref{fig:infra}.

\subsection{FPGA and ASIC Deployment}

\codename is built to generate vendor-agnostic synthesizable SystemVerilog RTL hardware components and necessary scripts to support both FPGA and ASIC implementation flows with various tools and targets.

\subsection{FPGA Toolflow}
\label{sec:fpga}

Our design decisions ensure that integrating the generated IP into an FPGA system is fairly straightforward.
Modern FPGAs are often Programmable System-on-Chips, where on-chip hardened processors and the FPGA fabric are interconnected to deploy heterogeneous systems.
We provide modular TCL scripts to enable such integration, which automates the following process.
On AMD-Xilinx FPGAs, the ZYNQ processing system is first configured to have three full AXI Subordinate ports with the maximum width (128-bit on ZCU104) for the highest throughput and one AXI-Lite Manager port for configuration.
Their address mappings are updated, and the IP ports are connected to them.
The TCL scripts then verify the connections, synthesize, place \& route the design, exporting bitstream and generating reports.

In addition to this standard flow, the standardized AXI interfaces of \codename allow the user to easily integrate the generated IP into existing signal processing dataflow systems.
Our scripts for AMD-Xilinx Vitis then create an application project, import the bitstream and C firmware, set compiler optimizations, and allow the user to execute the system as baremetal.

PetaLinux, the AMD-specific embedded Linux distribution, is widely used for higher-level applications with ZYNQ systems. 
When executed on PetaLinux, our C firmware manages the physical to virtual address mapping to provide seamless operation with an intuitive execution API, the same as baremetal.
Our C firmware is also wrapped in Python using ctypes, allowing users to pass and receive data as Numpy arrays in a Python environment.
This PYNQ API allows the user to be productive during prototyping, similar to the PYNQ API offered by \hlsml.
However, as the user moves towards production, our framework also allows layers of the same API to be extended into an optimized production-ready C runtime.

\subsection{ASIC Toolflow}
\label{sec:asic}

ASIC design is a process that often takes several months to years.
To obtain a first-order approximation of power-performance-area (PPA) analysis of a given CGRA configuration, we provide a basic set of TCL scripts for Cadence Genus \& Synopsys DesignCompiler for synthesis and Cadence Innovus for place \& route.
Users can link their PDKs, set a few basic parameters, and run the scripts to generate GDSII files and reports.
More experienced ASIC designers can further tweak the scripts for a more optimized design.
Some example GDSII outputs are shown in Fig.~\ref{fig:gds}.

We have partnered with ARM to tape out an SoC with the IP generated by \codename integrated into NanoSoC~\cite{nanosoc}, an open-source SoC platform around an ARM-M0 processor.
The platform features APB ports for configuration and three AHB Subordinate ports for data movement in its expansion region.
To make our design compatible with their system, we add AXI to AHB bridges around our design and a custom APB Subordinate port in place of the AXI-Lite port.

Together, the verification and deployment flows provided by \codename form a production-grade pipeline that supports model prototyping, debugging, and deployment in a unified environment. 
By enabling cycle-accurate testing with DPI-C integration, runtime configuration emulation, and full-stack simulation, the framework significantly reduces the time and risk associated with edge ML system development. 
Furthermore, the modular FPGA and ASIC toolchains allow seamless migration from early experimentation to large-scale integration or even silicon realization. 
Whether targeting a PYNQ-based research prototype or an industrial-grade SoC, \codename ensures that hardware, software, and machine learning models operate in concert, providing a reliable foundation for next-generation scientific computing at the edge.

\subsection{Ibex SoC Integration}

\revision{
To demonstrate end-to-end deployability on an open platform, we integrated cgra4ml with the lowRISC Ibex RISC-V SoC, which we build and simulate via the FuseSoC flow. 
We add an AXI crossbar to unify the engine’s three AXI managers into a single port, simplifying memory mapping and arbitration. 
We also add lightweight bridges that translate between the Ibex load/store interface and AXI, allowing the CGRA engine to present a device port for configuration and one host port to the system.
The subsystem is then packaged as a reusable FuseSoC IP core.
The hardware/firmware interface is platform agnostic.
On the software side, we compile our runtime and model firmware with a standard RISC-V toolchain, load the ELF together with quantized weights into on-chip SRAM, and execute the workload on the full SoC. 
This integration validates \codename’s hardware/firmware stack in a fully open, reproducible RISC-V environment and provides a template for broader open-source SoC deployments.
}
\section{Experimental Evaluation}
\label{sec:results}

In this section, we present the results of experiments comparing \codename to \hlsml on ZCU104 evaluation board with \emph{Zynq UltraScale+ XCZU7EV-FFVC1156-2-E MPSoC} FPGA. 
We also demonstrate \codename on deeper models such as ResNet-50 and PointNet (Table~\ref{tab:result_respoint}), which are too large to be implemented using \hlsml. Our ASIC results show the power and area efficiency of the RTL-based CGRA implemented using the ASIC flow provided with \codename.

\subsection{Performance Analysis and Practical Guidelines}
\label{sec:perf}
The performance of the CGRA for a given bundle can be analyzed based on the runtime parameters ($N,H,H_S,H_T,W,I,I_S,I_T,O,O_S,O_T,K_W,K_H$) described in Table~\ref{tab:runtime_params}.
The number of clock cycles and the off-chip data movement required to process a layer on a CGRA with $R,C$ rows \& columns of PEs are as follows:

\begin{align}
  \text{Clock cycles}  &= O_T I_T(1 + N H_T W (1 + I_S K_H)) \label{eq:clocks}  \\
  \text{Weight words}  &= O_T  I_T I_S  K_H  C  \\
  \text{Input words}   &= O_T I_T  N  H_T  W  I_S (R + K_H/2)  \\
  \text{Output words}  &= N H_{\textbf{out}} W_{\textbf{out}} O. \label{eq:out}
\end{align}

Increasing the number of PEs improves the performance by increasing the on-chip computation and reducing the number of iterations $O_T, H_T$. 
The ratio between peak performance ($RC{\times}\text{Frequency}$) and real performance (MAC operations in a layer/time) depends on the ratio of PEs that are idle when computing a layer as follows:
\begin{align}
  \text{Idle PE cols ratio}  
        &{=} [C\%K_W]/C{+}[O\%O_S]K_W/[CO_T] \label{eq:uncol} \\
  \text{Idle PE rows ratio}  
        &{=} [H{\%}R]/H   \label{eq:unrow}
\end{align}

Expression (\ref{eq:uncol}) is minimized by our hardware and unified dataflow design (Algorithm \ref{alg:dataflow}) that enables the $C$ columns of CGRA to dynamically group and process Out-Channels $O$. 
For example, for a CGRA with $C{=}96$, the ratio of idle columns in each iteration: $(C\%K_W)/C$ is zero for the most common layers with $K_W{=}\{1,3\}$, and $1\%, 5\%, 8\%$ for less common layers with $K_W{=}\{5,7,11\}$.
To further minimize the overall idle ratio of columns (Exp. \ref{eq:uncol}), the user can pick $C$ such that $\lfloor C/K_W \rfloor$ is a factor of $O$ for most layers. 
For ResNets, multiples of 3 and 4 ($C{=}12,24,96$), and for matrix multiplication based workloads, multiples of 8 ($C{=}8,16,32$) are such optimal choices. Among the $R$ columns of the CGRA, Expression \ref{eq:unrow} is minimized by choosing $R$ as the common factor of the $H$ of most layers of the DNNs the user wishes to run on the generated hardware. 
Figure \ref{fig:perfres} demonstrates the performance efficiency and off-chip data movement when processing ResNet50 on PE arrays of different sizes.

\begin{figure}
\centering
\begin{tikzpicture}[scale=0.80]
\definecolor{color1}{RGB}{13,161,177}
\definecolor{color2}{RGB}{176, 238, 245}
\definecolor{color3}{RGB}{241, 219, 204}
\definecolor{color4}{RGB}{244, 177, 131}
\definecolor{color5}{RGB}{197, 90, 16}
\definecolor{color6}{RGB}{16, 90, 197}
\begin{axis}[
    ybar stacked,
    ymin=0,
    ymax=180000,
    bar shift=-4pt,
    legend style={at={(0.2,-0.2)},
      anchor=north,legend columns=-1, nodes={scale=0.75, transform shape}},
    ylabel={Resource},
    symbolic x coords={12,24,48,72,96},
    xlabel={\# of PE columns},
    xtick=data,
    ]
\addplot +[fill=color1, draw=color1] coordinates {(12,20801) (24,31669) (48,52404) (72, 76891) (96,94578)};
\addplot +[fill=color2, draw=color2] coordinates {(12,4489) (24,8542) (48,18601) (72, 25245) (96,39391)};
\legend{LUTs (4 bit), LUTs (8 bit)}
\end{axis}
\begin{axis}[
    ybar stacked,
    ymin=0,
    ymax=180000,
    bar shift=8pt,
    legend style={at={(0.7,-0.2)},
      anchor=north,legend columns=-1, nodes={scale=0.75, transform shape}},
    symbolic x coords={12,24,48,72,96},
    xlabel={\# of PE columns},
    xtick=data,
    ]
\addplot +[fill=color4, draw=color4] coordinates {(12,36757) (24,50295) (48,77338) (72, 106168) (96,130129)};
\addplot +[fill=color3, draw=color3] coordinates {(12,2542) (24,4909) (48,8359) (72, 14488) (96,19525)};
\legend{FFs (4 bit) , FFs (8 bit) }
\end{axis}

\begin{axis}[
  ymin=0,
  ymax=3,
  ylabel near ticks, 
  yticklabel pos=right,
  ylabel={Power},
  symbolic x coords={12,24,48,72,96},
  legend style={nodes={scale=0.7, transform shape}},
  ]
\addplot+[smooth, color=color5,
            mark=*,
            mark options={fill=color5},] 
    coordinates {(12,1.063) (24,1.175) (48,1.567) (72,1.798) (96,1.801)};
\addplot+[smooth, color=color6,
            mark=*,
            mark options={fill=color6},]
    coordinates {(12,1.156) (24,1.242) (48,1.713) (72,1.872) (96,2.197)};
\legend{Total PL power (4 bit) , Total PL power (8 bit) }      
\end{axis}

\end{tikzpicture}
\Description{Scaling of hardware resource and power consumption}
\caption{Scaling of hardware resource and power consumption on a Xilinx ZCU104 FPGA, for different bitwidths and PE columns, with PE rows = 8}
\label{fig_resource}
\end{figure}
\begin{figure}
    \centering
    \includegraphics[width=0.8\linewidth]{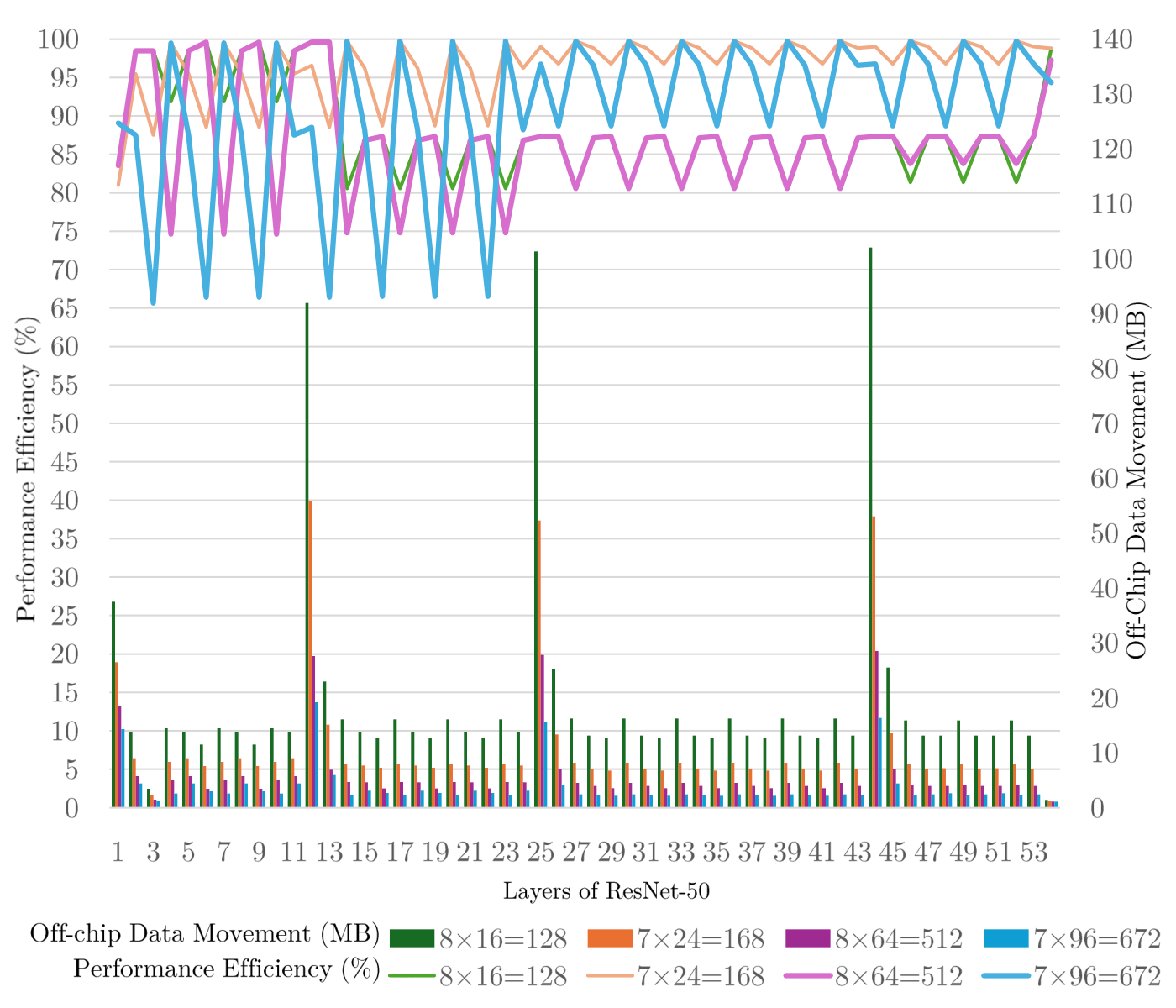}
    \Description{Performance analysis of ResNet-50}
    \caption{
Performance analysis of ResNet-50 on different CGRA array sizes. 
Performance efficiency measures how well each layer is dynamically mapped over statically configured hardware.
While larger arrays reduce off-chip data movement due to high data reuse, they have lower utilization for smaller layers.
The unified dataflow of the array (Sec.~\ref{subsec:reconfig}) maximizes the utilization to near 100\% and minimizes data movement for most layers.
}
\label{fig:perfres}
\end{figure}

\begin{table}
    \newcommand{\mc}{\multicolumn}
    \newcommand{\cgr}{\codename}
    \newcommand{\mct}{\mc{2}{c}}
    \newcommand{\cmdtt}{\cmidrule(l){2-3} \cmidrule(l){4-5} \cmidrule(l){6-6} \cmidrule(l){7-8} \cmidrule(l){9-10} \cmidrule(l){11-11}}
    \newcommand{\cmdf}{\cmidrule(l){2-6} \cmidrule(l){7-11}}
    \newcommand{\tvm}{TVM}
    \renewcommand{\arraystretch}{1.2}
    \centering
    \begin{tabular}{rccc} 
    \toprule
     Model                   & ResNet-50        & PointNet~\cite{pointnet}     & Autoencoder~\cite{mlperf}  \\ \midrule
     Bits                    & 4                & 4            & 8\\ 
     Operations              & 5.36 billion     & 73.7 million & 2.09 million    \\ \midrule
     PEs                     & (7,96)           & (32,32)     & (16, 64)\\
     Frequency (MHz)         & 250              &  250       & 250\\
     FFs                     & 101706           & 69277     & 154247 \\
     LUTs                    &  82200           & 100076    & 140357 \\ 
     BRAMs                   & 6                & 4.5       & 4.5   \\ \midrule
     Static Power (W)        &   0.700          & 0.700     & 0.702  \\ 
     Dynamic Power (W)       & 3.847            &  3.840    & 4.196  \\
     Total Power (W)         & 4.547            &  4.540    & 4.898  \\ 
     GOPs/W                  & 37.3             &  56.8     & 1\zhenghua{What is this?}  \\ 
     \bottomrule
    \end{tabular}
    \vspace{5pt}
    \caption{Implementation of ResNet-50, Pointnet and Autoencoder on ZCU104 FPGA}
    \label{tab:result_respoint}
\end{table}

Fig. \ref{fig_resource} demonstrates the resource and power consumption for different PE arrays on the FPGA, with rows=8. Since larger arrays are more efficient in terms of GOPs/W, users should strive to implement the largest possible array within the available resources.

\begin{table}[h]
\centering
\begin{tabular}{l r}
\hline
\textbf{Operation} & \textbf{Percentage} \\
\hline
Writeback pointer calculations & 47.37\% \\
Writeback                      & 24.34\% \\
Cache flush write              & 2.14\%  \\
Add IT passes                  & 1.10\%  \\
Add Bias                       & 4.50\%  \\
Activation                     & 6.04\%  \\
Residual add                   & 7.50\%  \\
Pooling                        & 4.94\%  \\
Handshake                      & 1.18\%  \\
\hline
\end{tabular}
\caption{Breakdown of performance penalty of the CPU workload for ResNet-50}
\label{tab:cpuperf}
\end{table}

\subsection{Performance Evaluation of CPU operations}
\label{sec:perfcpu}

\revision{
We partition the workload between the CPU and the CGRA as shown in Fig.~\ref{fig:bundle}.
Lightweight, pointwise operations are done in the CPU, while heavy, parallelizable operations are done through our CGRA.
While this partition allows us to implement additional pointwise operations and activations as required, with minimal modification of hardware, our evaluations show that often CPU operations become the bottleneck in addition to data transfer between the CPU and the custom hardware.
Depending on the workload, over 50\% of the execution time is spent on CPU operations, and Table~\ref{tab:cpuperf} demonstrates the breakdown of execution time in the CPU for ResNet-50. 
As observed, we can achieve significant speedup by partially parallelizing the most compute-intensive operations, such as calculating writeback pointers and packing writeback words on custom hardware.
We are currently working on this to achieve the performance boost without sacrificing extensibility of the bundle architecture.
Our fusion strategy includes writing an extensive set of bundle parameters to a register bank at startup, breaking the CGRA-CPU dual execution into a three-stage flow of CGRA-CPU-Writeback with locks and handshaking for synchronization. 
The new writeback stage would increase the performance by packing output bits, calculating all pointers in parallel and performing writes without CPU intervention.
}

\subsection{Micro-benchmarking}
\begin{figure}
    \centering
    \includegraphics[width=0.9\linewidth]{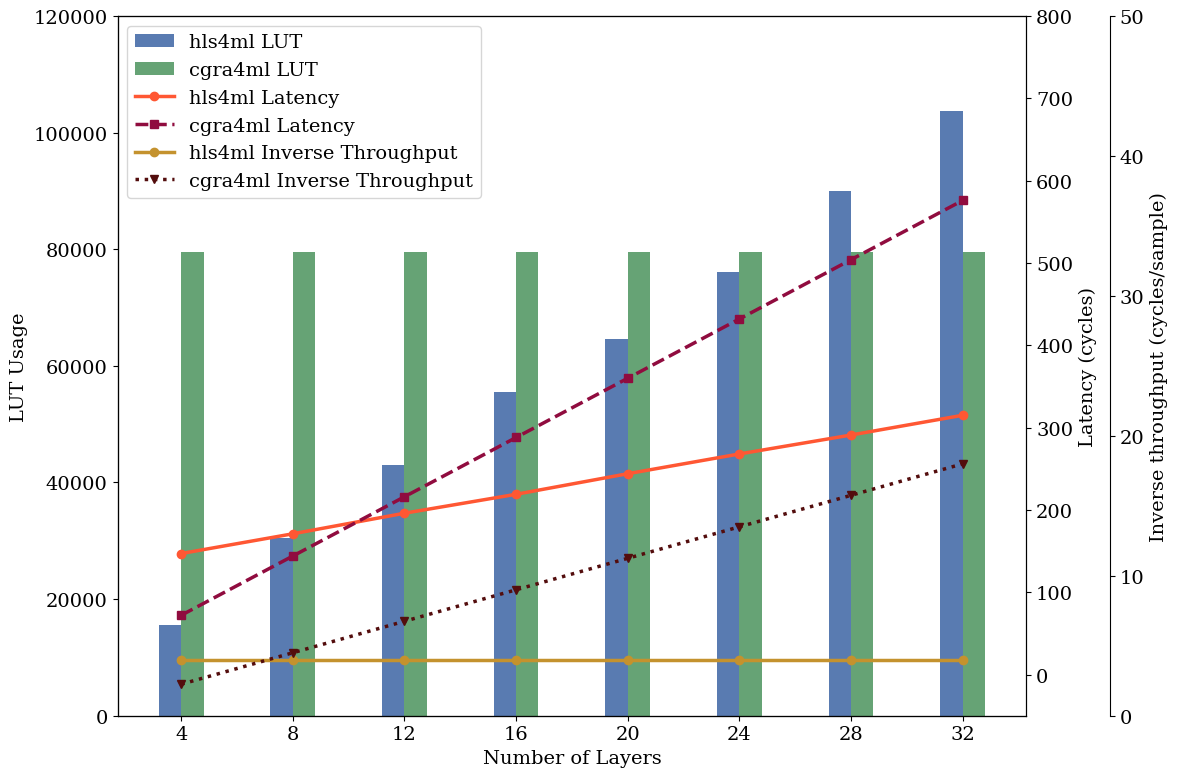}
    \Description{Synthetic workloads of cascading dense layers}
    \caption{Synthetic workloads of \revision{dense autoencoder models with 4-32 layers (x-axis), where each dense layer has 16 inputs and 16 outputs and ReLU activation is used between the layers. The model is quantized to have 8-bit inputs, weights, and biases, which are randomized with a normal distribution.} Inference is performed with a batch size of 32. \revision{The CGRA is configured to have 16 rows, 32 columns, an accumulator width of 24 bits, and 128-bit wide AXI ports. \hlsml stores weights on-chip due to its dataflow architecture, while \codename stores weights off-chip, but buffers them in ping-pong buffers to reuse across the inference of a layer. We do the benchmarking on the ZCU104 development board. } \codename maintains flat resource usage, while \hlsml-generated hardware shows linear growth with the number of layers. We report steady-state inverse-throughput, which is the II (initiation interval) for \hlsml. Latency increases linearly for both, with \hlsml showing a slower rate. }
    \label{fig:microbenchmarking}
\end{figure}

We first conduct a micro-benchmarking experiment to observe how the resource usage and latency of hardware generated by \hlsml and \codename scale with neural network size. The synthetic workload is a stack of dense layers with ReLU between layers; each layer is $16{\times}16$ with bias. All tensors (inputs, weights, biases) are 8-bit quantized; weights and biases are initialized from a zero-mean normal distribution. We report the time to process one batch of 32 inputs. The \codename CGRA uses a $16{\times}32$ PE array with 24-bit accumulators and communicates via 128-bit AXI. On ZCU104, \hlsml stores per-layer weights on-chip (consistent with its layer-specialized dataflow), whereas \codename streams weights from off-chip DDR but employs double (ping–pong) buffering to maximize reuse within each layer. Inputs/outputs are transferred over 128-bit DMA, and layer outputs are requantized to 8-bit with round-to-nearest and saturation. The results are summarized in Fig.~\ref{fig:microbenchmarking}.

The observations align with our discussion in Section~\ref{sec:background}. \hlsml, a representative example of layer-by-layer hardware generation, produces designs whose resource usage grows linearly with the number of layers. The hardware dataflow is well pipelined, leading to relatively modest growth in latency. In contrast, \codename is a CGRA-based accelerator that reuses the same core across layers, resulting in flat resource consumption. The latency still grows linearly, but limited pipelining and additional dataflow-control overhead lead to a steeper slope. Based on these trends, we identify three regimes by network depth: for small networks (4–8 layers), \codename achieves lower latency (at higher resource usage) due to its high-throughput systolic array; for mid-sized networks (12–24 layers), \hlsml provides both lower latency and resource usage; beyond 28 layers, \codename becomes attractive because of its constant resource footprint, provided the application can tolerate the higher latency relative to layer-specialized designs.

\begin{figure}
    \centering
    \includegraphics[width=0.8\linewidth]{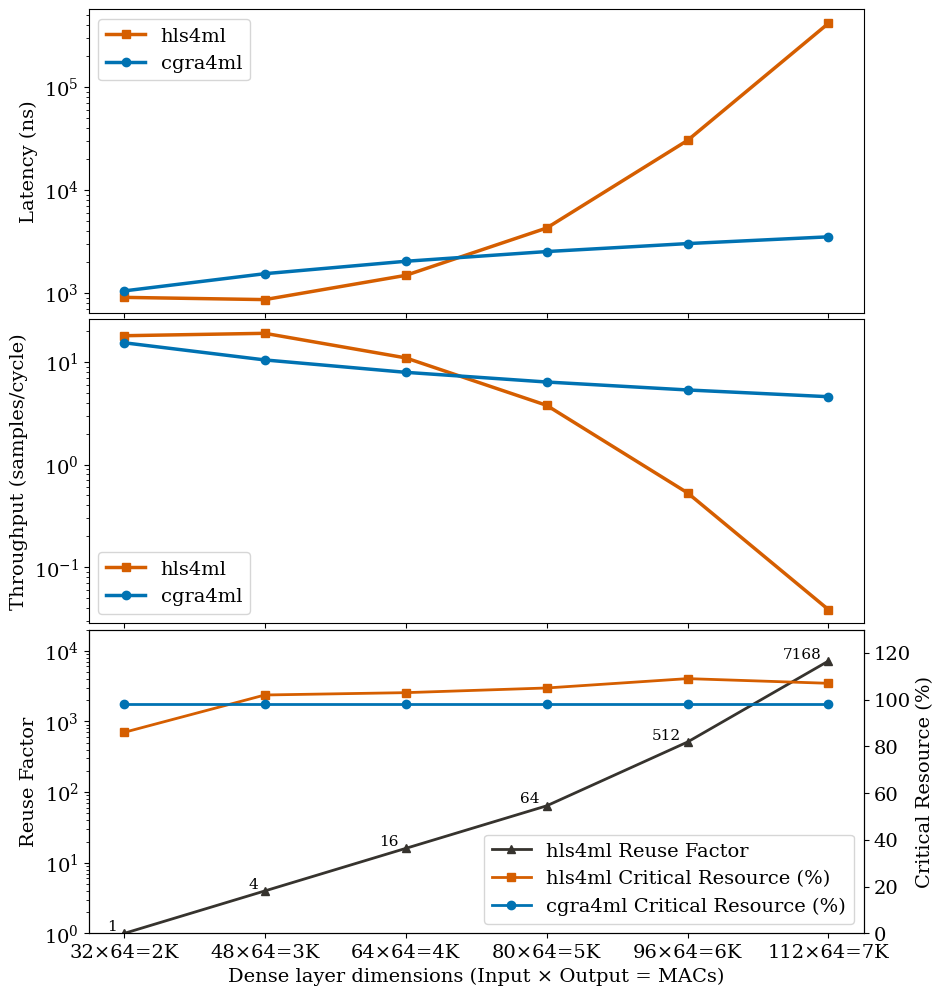}
    \caption{\textbf{Iso-Resource Microbenchmarking} on Pynq-Z2 at clock frequency of 150 MHz and batch size of 16. For \codename, we find a 16×16 PE array (8×8 MAC, 24-bit accumulator) fills the FPGA.  For \hlsml, we find the largest dense layer (8×8 MAC) that can fit within the FPGA with no resources exceeding 100\% is that with 32 inputs and 64 outputs (2048 multiply-accumulate operations) at reuse factor of 1. We then increase the workload size linearly by 1024 MACs, and find the reuse factor needed to fit that workload in the FPGA using \hlsml. We find that the reuse factor needs to be increased almost exponentially for linear increase in workload size, causing latency and throughput to degrade for large workload sizes.}
    \Description{A comparison of Iso-Resource Microbenchmarking on a Pynq-Z2 FPGA at 150 MHz. The data shows that while a cgra4ml 16×16 PE array utilizes the full FPGA resources efficiently, hls4ml requires an exponential increase in the reuse factor to accommodate linear increases in workload size (starting from a 32x64 dense layer). This scaling limitation in hls4ml leads to a significant degradation in latency and throughput for larger workloads compared to the CGRA approach.}
    \label{fig:isolation}
\end{figure}

\revision{
To complement Fig.~\ref{fig:microbenchmarking}, we perform an iso-resource benchmarking on a Pynq-Z2 (150 MHz) with 8-bit activations/weights. 
For \hlsml, we start with a dense layer of 32 inputs and 64 outputs, which we find is the largest dense layer that fits on the FPGA at reuse factor of 1 without any resources exceeding 100\%.
The workload size is $32{\times}64{=}2048$ multiply-accumulate operations.
We then increase the workload size linearly by 1024, finding the reuse factor necessary to fit it into the FPGA.
For CGRA4ML, we select the largest array that fits, which is 16×16 PEs with 8-bit × 8-bit multiply and 24-bit accumulation.
In our experiment, we find that LUTs are the bottleneck resource. 
We run them at the same frequency of 150 MHz and batch size of 16. 
This way, we keep the resource utilization almost constant (100\%) and same for \codename and \hlsml throughout the experiment.
We report steady-state batch latency and throughput for both designs in  Fig.~\ref{fig:isolation} for every workload size.
}

\revision{
Increasing \hlsml reuse is often effective for fitting larger layers.
But we find that to fit linearly larger workload size (number of multiply-accumulates), the reuse factor needs to be increased exponentially as shown in Fig.~\ref{fig:isolation}.
This causes latency and throughput to degrade similarly.
At high reuse such as 512, \hlsml latency exceeds that of CGRA4ML, which maintains stable latency and higher throughput at the matched resource point. 
We also found for input size=128 and output size=64, we could not get resource utilization close to 100\% even with reuse factor of 8192, which is the largest possible reuse factor for a dense layer with 8192 MAC operations.
Therefore, we suggest the users to prefer hls4ml when the model fits with low reuse factor on the target device, and use CGRA4ML when extreme reuse is required to meet resource limits.
}



\subsection{ResNet-50}

ResNet-50 \cite{resnet} is a deep convolutional neural network with 5.36 billion operations in 53 convolutional layers, one dense layer, two pooling layers, and 17 skip connections. Its input is a $224{\times}224{\times}3$ tensor, and outputs are 1000 classes, and it is run on an array of $7{\times}96{=}672$ PEs.

Performance analysis of ResNet-50 on different CGRA array sizes is shown in Fig.~\ref{fig:perfres}. 
Performance Efficiency is the ratio of (\#PEs${\times}$clocks)/(MACs needed for the layer). 
It indicates the utilization of the MAC units over space and across time, and quantifies how well layers of different shapes and nature are dynamically mapped to the statically configured PE array.
As discussed in Section \ref{sec:perf}, the unified dataflow with dynamic reconfiguration maximizes efficiency and minimizes off-chip data movement via heavy data reuse across the layers.
Per expressions \ref{eq:uncol} \& \ref{eq:unrow}, array sizes of $7{\times}24$ and $7{\times}96$ provide high utilization for ResNet-50, since its input height ($H{=}224$) is a multiple of 7, and the out-channels $O$ in computation-heavy 3x3 conv layers are multiples of 4.
Off-chip data movement includes inputs, weights, and outputs. 
While larger arrays have lower data movement due to higher data reuse, they are underutilized for $1{\times}1$ layers with out-channels fewer than PE cols.
The spikes in data movement correspond to $1{\times}1$ layers with 2048 out-channels, where weights are only reused across the PE array.

\begin{table}
    \newcommand{\mc}{\multicolumn}
    \newcommand{\cgr}{\codename}
    \newcommand{\mct}{\mc{2}{c}}
    \newcommand{\cmdtt}{\cmidrule(l){2-3} \cmidrule(l){4-6}}
    \newcommand{\cmdf}{\cmidrule(l){2-6} \cmidrule(l){7-11}}
    \newcommand{\tvm}{TVM}
    \renewcommand{\arraystretch}{1.2}
    \centering
    \begin{tabular}{rccccc} 
    \toprule
     Framework         & \mc{2}{c}{\codename}    & \mc{3}{c}{\hlsml}               \\ \cmdtt
     Reuse factor      & -         & -           & 1        & 16       & 64        \\ 
     PEs               & (8,12)   & (16,32)      & -        & -        & -         \\
     Frequency (MHz)   & 250       & 250         &  250     &  250     &  250      \\ \cmdtt
     FFs               & 32451    & 64066      & 21428    & 20991    & 21335     \\
     LUTs              & 17959    & 43320      & 11810   & 11986   & 11963    \\ 
     BRAMs             & 4.5         & 4.5        & 5      & 5      & 7       \\ \cmdtt
     Static Power (W)  & 0.694     & 0.697       & 0.693    & 0.693    & 0.693     \\ 
     Dynamic Power (W) & 3.078     & 3.450       &  2.948   &  2.886   & 2.820    \\
     Total Power (W)   & 3.772     & 4.147       &  3.642   &  3.579   & 3.513    \\ 
     \bottomrule
    \end{tabular}
    \vspace{5pt}
    \caption{Comparing Jet-tagger model on \codename with different array sizes and on \hlsml with varying reuse factors}
    \label{tab:fpga}
\end{table}

\subsection{PointNet \& Jet Tagger}


Jet Tagger is a small model with 117K operations in 4 dense layers, widely used for benchmarking. 
Table~\ref{tab:fpga} compares the Jet Tagger model implemented using \codename with two different CGRA configurations (8×12 and 16×32 PEs) against \hlsml across reuse factors of 1, 16, and 64. 
Notably, \hlsml achieves lower overall resource usage (FFs and LUTs) and power consumption for this small model. 
This is expected, as Jet Tagger’s modest compute demand and regular layer structure are well-suited to \hlsml’s layer-by-layer, statically scheduled architecture, which minimizes hardware overhead when the model fits entirely within the available logic. 
In contrast, \codename incurs additional overhead from its reconfigurable CGRA infrastructure, runtime firmware, and AXI interfacing, which are optimized for larger, more complex models. 
For tiny workloads like Jet Tagger, this overhead outweighs the performance benefits of dynamic reconfiguration. 
Thus, while \codename provides a scalable solution for deeper or irregular networks, \hlsml remains more resource- and power-efficient for simple, compact models, highlighting a natural tradeoff between generality and specialization.

PointNet is a model designed for particle classification.
It includes three 1D convolutional layers, three dense layers, and two pooling layers.
This results in approximately 1.18 billion operations for a batch size of 16.
The model does not synthesize with \hlsml due to its depth, irregular structure, and extensive activation reuse.
\hlsml’s layer-by-layer pipeline architecture requires each layer to be synthesized as a standalone datapath.
The entire model must fit into the FPGA fabric simultaneously.
Even with reuse factor tuning, \hlsml fails to compile PointNet on the ZCU104 due to resource exhaustion during synthesis and placement.
In contrast, \codename supports PointNet by reusing its CGRA fabric across all layers.
This enables not only functional deployment but also high energy efficiency (56.8 GOPs/W), making \codename uniquely capable of supporting such mid-sized scientific neural networks that exceed the scalability bounds of traditional HLS-based approaches.



\subsection{ASIC Implementation}

\begin{table}
    \setlength{\tabcolsep}{15pt}
    \renewcommand{\arraystretch}{1.2}
    \centering
    \begin{tabular}{rcccc} 
    \toprule
     Technology         & \multicolumn{1}{c }{7nm} & \multicolumn{1}{c }{28nm}  & \multicolumn{1}{c}{65nm}  \\ 
     \midrule
     Frequency (MHz)    &     500        & 500      &  250                      \\
     Area (mm$^2$)      &     0.037      & 0.175    &  0.502                    \\
     Static power (mW)  &     0.614      & 25.58    &  0.130                    \\
     Dynamic power (mW) &     29.8       & 46.87    &  80.71                    \\
     Total power (mW)   &     30.4       & 72.45    &  80.84                    \\ 
     GOPs/W             &     3158       & 1325     &  594                      \\ 
     \bottomrule
    \end{tabular}
    \vspace{5pt}
    \caption{Power/Performance/Area (PPA) analysis across technology nodes for a CGRA configuration of $R{\times}C=8{\times}24=192$. 28 \& 65\,nm results use a Cadence Genus/Innovus flow; 7\,nm results use a Synopsys DC/ICC2 flow}
    \label{tab:asic}
\end{table}







\newlength{\imgsep}    \setlength{\imgsep}{4mm}
\newlength{\rowwidth}  \setlength{\rowwidth}{0.8\linewidth}

\newlength{\pairwA}    
\newlength{\pairwB}    
\setlength{\pairwA}{0.5\rowwidth}                 
\setlength{\pairwB}{\dimexpr \rowwidth-\imgsep-\pairwA\relax}

\begin{figure}[t]
  \centering

  \begin{subfigure}[t]{\rowwidth}
    \centering
    \includegraphics[width=\linewidth]{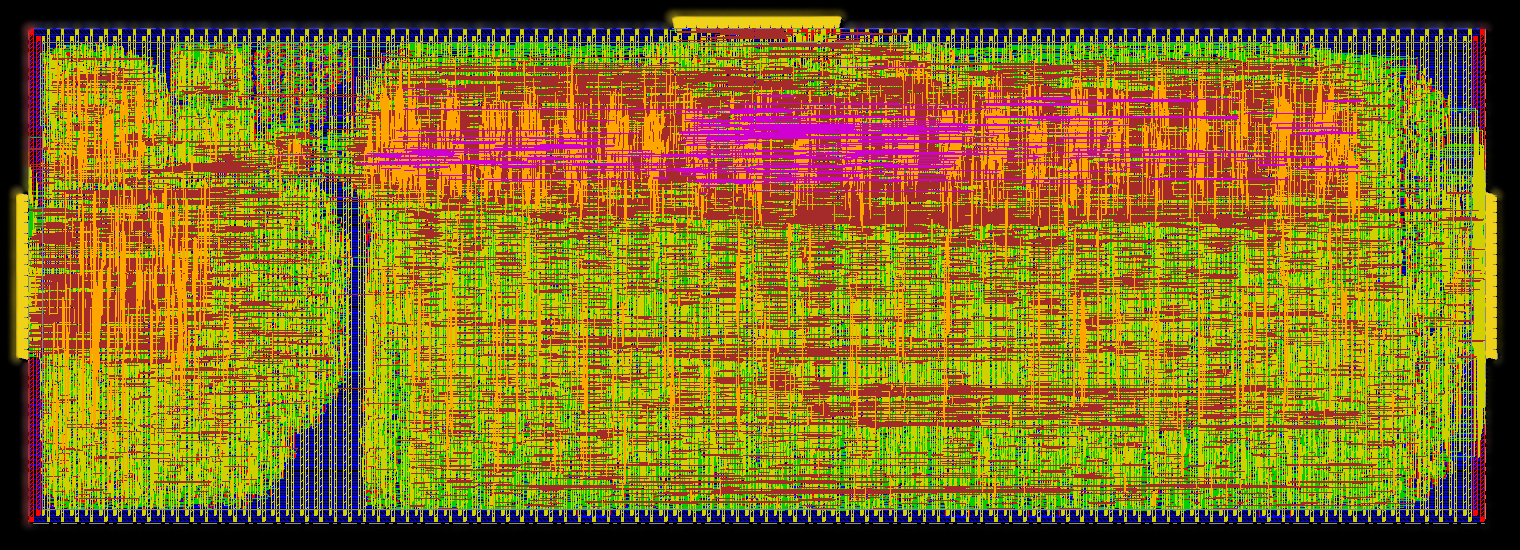}
    \caption{TSMC 65nm LP (low power)}
  \end{subfigure}

  \vspace{\imgsep}

  \begin{subfigure}[t]{\pairwA}
    \centering
    \includegraphics[width=\linewidth]{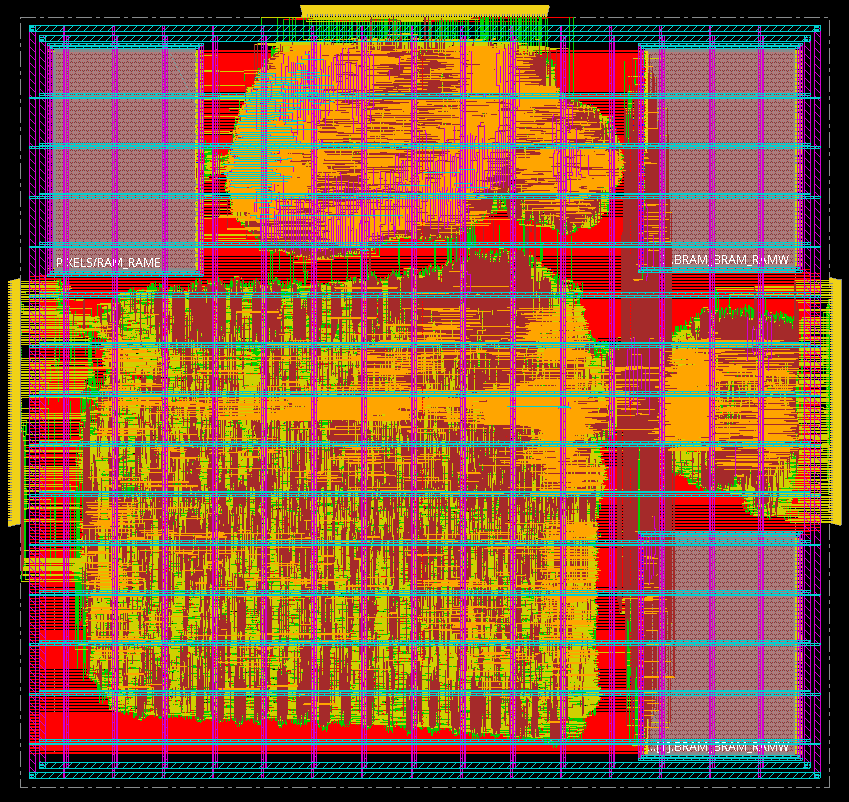}
    \caption{TSMC 28nm HPC+ (high performance)}
  \end{subfigure}\hspace{\imgsep}%
  \begin{subfigure}[t]{\pairwB}
    \centering
    \includegraphics[width=\linewidth]{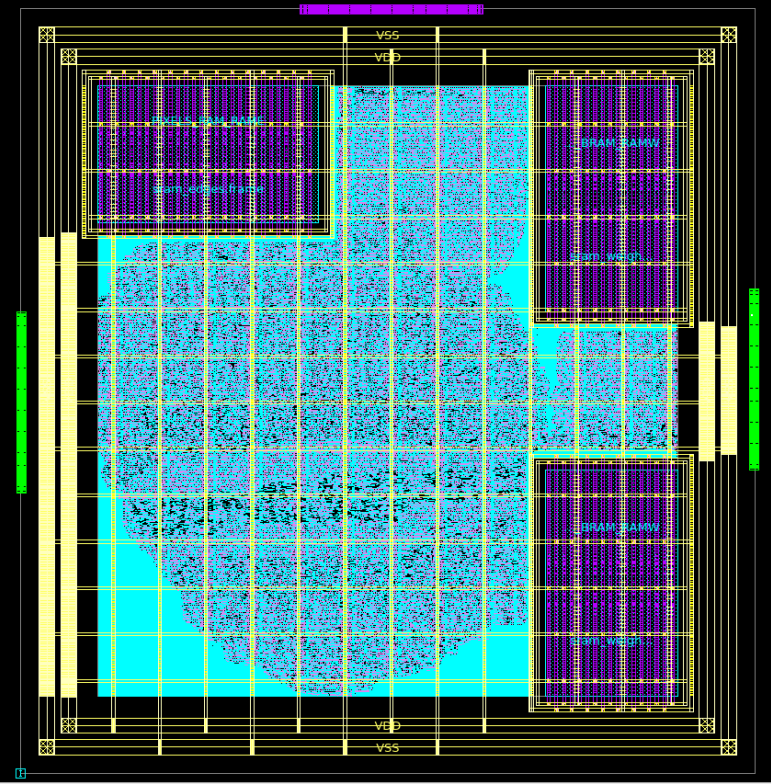}
    \caption{TSMC 7nm}
  \end{subfigure}

  \Description{(a) 65nm view. (b) 28nm view. (c) 7nm view.}
  \caption{CGRA of $R{\times}C{=}8{\times}24$ configuration generated by \codename and taken through the preliminary ASIC scripts. (a) and (b) are synthesized with Cadence Genus and placed \& routed with Cadence Innovus, while (c) was synthesized with Synopsys Design Compiler and placed \& routed using Synopsys ICC2.}
  \label{fig:gds}
\end{figure}

\revision{
As described in Sec.~\ref{sec:asic}, we built a complete system-on-chip (SoC) around the \codename CGRA using NanoSoC~\cite{nanosoc}, an open-source ARM-based SoC generation platform developed with ARM Research. 
The CGRA connects to the ARM Cortex-M subsystem through DMA360, which bridges the accelerator’s three AXI-Stream ports to the AHB interconnect for programmable, high-throughput transfers. 
Macros for on-chip memories (weights/pixels) are generated with ARM Artisan Memory Compiler, and the SoC is synthesized in Cadence Genus and placed \& routed in Cadence Innovus targeting TSMC 28\,nm HPCPLUS, as shown in Fig.~\ref{fig:gds}. 
This validates that \codename emits silicon-ready RTL that integrates cleanly into a production-style SoC environment.
}

\revision{
Beyond first-order synthesis, we executed place-and-route flows across multiple nodes:
\begin{enumerate}
    \item TSMC 65\,nm LP with Cadence Genus/Innovus
    \item TSMC 28\,nm HPCPLUS with Cadence Genus/Innovus, and
    \item TSMC 7\,nm with Synopsys DC/ICC2
\end{enumerate}
For 65\,nm, we targeted 250\,MHz with a custom floorplan that partitions the weight memory into narrow (8-bit) SRAM macros placed symmetrically near AXI-Stream port for weights, and a 12-bit macro for pixels; the resulting design uses 9 SRAM macros.
At 28\,nm and 7\,nm, the SRAM instances are sufficiently compact to avoid width-splitting; these designs employ 3 macros total.
Across nodes we used scripted flows from synthesis through P\&R, including memory macro integration, floorplanning, clocking, routing, and iterative DRC cleanup.
}

\revision{
In addition to the 65\,nm and 28\,nm implementations above, we evaluated larger \codename-generated arrays in TSMC 7\,nm. 
The 192-PE configuration closes at 1\,GHz with sub-0.5\,W total power (Table~\ref{tab:asic}), illustrating the scalability of \codename toward high-throughput ASIC targets. 
By varying $R{\times}C$ and memory organization from the Python frontend, users can rapidly explore power, performance, and area trade-offs across process nodes and CGRA sizes.
}

These results validate that the RTL emitted by \codename is not tied to any single foundry or node, but can be retargeted across a wide spectrum of technologies.
The ability to support both bleeding-edge 7nm and legacy 65nm nodes makes \codename suitable for deployment in academic tapeouts, low-cost SoCs, and commercial research settings alike.
By varying the CGRA configuration parameters from the Python frontend, users can easily explore the power, performance, and area trade-offs across process technologies.
This highlights the practical utility of \codename not only as a research framework but also as a prototyping and pre-silicon exploration tool for real ASIC targets.
\revision{
The scripts and toolflows we provide are for preliminary exploration.
Due to their nature each tapeout requires an experienced physical designer to extensively tweak the scripts and flow to generate results free of DRC violations. 
}
\codename allows rapid iteration on ASIC-ready CGRA accelerators, from low-power edge systems to high-performance scientific computing chips.

\section{Future Work}

We are expanding the framework's capabilities to support more complex neural network architectures, including transformers, to broaden its applicability in scientific edge computing further.
\revision{We are also working on migrating more and more runtime components to hardware, to parallelize their computation and minimize any CPU bottlenecks.}
\section{Conclusion}

In this paper, we introduced \codename, a modular framework for deploying large-scale neural networks in scientific edge computing. 
By supporting off-chip storage, diverse model architectures, and holistic ASIC/FPGA flows, \codename complements existing tools like \hlsml while enabling more complex and performant designs. 
Its vendor-agnostic RTL generation, runtime firmware, and integrated verification infrastructure reduce development overhead and simplify hardware-software co-design. 
We believe \codename offers a practical and extensible foundation for advancing high-performance scientific computing at the edge.

\bibliographystyle{ACM-Reference-Format}
\bibliography{ref}

@misc{mlperf,
  author       = {TinyML Team},
  title        = {MLPerf™ Tiny Deep Learning Benchmarks for Embedded Devices},
  year         = 2025,
  url          = {https://github.com/mlcommons/tiny}
}

@misc{torchdynamo,
  author       = {PyTorch Team},
  title        = {PyTorch Dynamo},
  year         = 2025,
  url          = {https://docs.pytorch.org/docs/stable/torch.compiler_dynamo_overview.html}
}

@misc{openaitriton,
  author       = {OpenAI Team},
  title        = {OpenAI Triton},
  year         = 2025,
  url          = {https://openai.com/index/triton/}
}

@INPROCEEDINGS{VecPAC,
  author={Tan, Cheng and Patil, Deepak and Tumeo, Antonino and Weisz, Gabriel and Reinhardt, Steve and Zhang, Jeff},
  booktitle={2023 IEEE/ACM International Conference on Computer Aided Design (ICCAD)}, 
  title={VecPAC: A Vectorizable and Precision-Aware CGRA}, 
  year={2023},
  volume={},
  number={},
  pages={1-9},
  doi={10.1109/ICCAD57390.2023.10323910},
  ISSN={1558-2434},
  month={Oct},}

@inbook{MLCGRA,
author = {Luo, Yixuan and Tan, Cheng and Agostini, Nicolas Bohm and Li, Ang and Tumeo, Antonino and Dave, Nirav and Geng, Tong},
title = {ML-CGRA: An Integrated Compilation Framework to Enable Efficient Machine Learning Acceleration on CGRAs},
year = {2025},
isbn = {9798350323481},
publisher = {IEEE Press},
url = {https://doi.org/10.1109/DAC56929.2023.10247873},
booktitle = {Proceedings of the 60th Annual ACM/IEEE Design Automation Conference},
pages = {1–6},
numpages = {6}
}

@INPROCEEDINGS{opencgra,
  author={Tan, Cheng and Xie, Chenhao and Li, Ang and Barker, Kevin J. and Tumeo, Antonino},
  booktitle={2020 IEEE 38th International Conference on Computer Design (ICCD)}, 
  title={OpenCGRA: An Open-Source Unified Framework for Modeling, Testing, and Evaluating CGRAs}, 
  year={2020},
  volume={},
  number={},
  pages={381-388},
  doi={10.1109/ICCD50377.2020.00070}}

@INPROCEEDINGS{aurora,
  author={Tan, Cheng and Xie, Chenhao and Li, Ang and Barker, Kevin J. and Tumeo, Antonino},
  booktitle={2021 Design, Automation \& Test in Europe Conference \& Exhibition (DATE)}, 
  title={AURORA: Automated Refinement of Coarse-Grained Reconfigurable Accelerators}, 
  year={2021},
  volume={},
  number={},
  pages={1388-1393},
  doi={10.23919/DATE51398.2021.9473955}}

@inproceedings{picachu,
author = {Qin, Jiajun and Xia, Tianhua and Tan, Cheng and Zhang, Jeff and Zhang, Sai Qian},
title = {PICACHU: Plug-In CGRA Handling Upcoming Nonlinear Operations in LLMs},
year = {2025},
isbn = {9798400710797},
publisher = {Association for Computing Machinery},
address = {New York, NY, USA},
url = {https://doi.org/10.1145/3676641.3716013},
doi = {10.1145/3676641.3716013},
booktitle = {Proceedings of the 30th ACM International Conference on Architectural Support for Programming Languages and Operating Systems, Volume 2},
pages = {845–861},
numpages = {17},
location = {Rotterdam, Netherlands},
series = {ASPLOS '25}
}

@article{autoqkeras,
  title={Automatic heterogeneous quantization of deep neural networks for low-latency inference on the edge for particle detectors},
  author={Coelho, Claudionor N and Kuusela, Aki and Li, Shan and Zhuang, Hao and Ngadiuba, Jennifer and Aarrestad, Thea Klaeboe and Loncar, Vladimir and Pierini, Maurizio and Pol, Adrian Alan and Summers, Sioni},
  journal={Nature Machine Intelligence},
  volume={3},
  number={8},
  pages={675--686},
  year={2021},
  publisher={Nature Publishing Group}
}

@misc{hls4ml,
  author       = {{FastML Team}},
  title        = {fastmachinelearning/hls4ml},
  year         = 2024,
  publisher    = {Zenodo},
  version      = {v0.8.1},
  doi          = {10.5281/zenodo.1201549},
  url          = {https://github.com/fastmachinelearning/hls4ml}
}

@misc{nanosoc,
  author       = {David Flynn, John Darlington, Daniel Newbrook},
  title        = {Nanosoc re-usable MCU platform},
  year         = 2023,
  url          = {https://soclabs.org/project/nanosoc-re-usable-mcu-platform}
}

@article{flexcnn,
    author = {Basalama, Suhail and Sohrabizadeh, Atefeh and Wang, Jie and Guo, Licheng and Cong, Jason},
    title = {FlexCNN: An End-to-end Framework for Composing CNN Accelerators on FPGA},
    journal = {ACM Transactions on Reconfigurable Technology and Systems},
    year = {2023},
    issue_date = {June 2023},
    publisher = {Association for Computing Machinery},
    address = {New York, NY, USA},
    volume = {16},
    number = {2},
    issn = {1936-7406},
    url = {https://doi.org/10.1145/3570928},
    doi = {10.1145/3570928},
    month = {mar},
    articleno = {23},
    numpages = {32}
}

@misc{vitisai,
  author       = {{AMD-Xilinx}},
  title        = {Vitis AI},
  year         = 2024,
  url          = {https://www.xilinx.com/products/design-tools/vitis/vitis-ai.html}
}

@misc{openvino,
  author       = {{Intel}},
  title        = {OpenVINO™ toolkit},
  year         = 2024,
  url          = {https://www.intel.com/content/www/us/en/developer/tools/openvino-toolkit/overview.html}
}

@misc{dlp,
  author       = {{Mathworks}},
  title        = {Deep Learning Processor IP Core Generation for Custom Board},
  year         = 2024,
  url          = {https://www.mathworks.com/help/deep-learning-hdl/ug/define-custom-board-and-reference-design-for-dl-ip-core-workflow.html}
}

@misc{qkeras,
  author       = {{QKeras Team}},
  title        = {QKeras},
  year         = 2024,
  version      = {v0.9},
  url          = {https://github.com/google/qkeras}
}

@misc{alexaxi,
  author       = {{Alex Forencich}},
  title        = {Verilog AXI Components},
  year         = 2024,
  url          = {https://github.com/alexforencich/verilog-axi/}
}

@misc{alexaxistream,
  author       = {{Alex Forencich.}},
  title        = {Verilog AXI Stream Components},
  year         = 2024,
  url          = {https://github.com/alexforencich/verilog-axis/}
}

@inproceedings{part2022,
  title={Particle transformer for jet tagging},
  author={Qu, Huilin and Li, Congqiao and Qian, Sitian},
  booktitle={International Conference on Machine Learning},
  pages={18281--18292},
  year={2022},
  organization={PMLR}
}

@INPROCEEDINGS{surveyOfTools,
  author={Rahimifar, Mohammad Mehdi and Granger, Charles-Etienne and Wingering, Quentin and Gouin-Ferland, Berthié and Rahali, Hamza Ezzaoui and Corbeil Therrien, Audrey},
  booktitle={2022 IEEE Nuclear Science Symposium and Medical Imaging Conference (NSS/MIC)}, 
  title={A Survey of Machine Learning to FPGA Tool-Flows for Instrumentation}, 
  year={2022},
  volume={},
  number={},
  pages={1-4},
  doi={10.1109/NSS/MIC44845.2022.10399017}}

@article{AIgean,
author = {Tarafdar, Naif and Di Guglielmo, Giuseppe and Harris, Philip C. and Krupa, Jeffrey D. and Loncar, Vladimir and Rankin, Dylan S. and Tran, Nhan and Wu, Zhenbin and Shen, Qianfeng and Chow, Paul},
title = {AIgean: An Open Framework for Deploying Machine Learning on Heterogeneous Clusters},
year = {2022},
issue_date = {September 2022},
publisher = {Association for Computing Machinery},
address = {New York, NY, USA},
volume = {15},
number = {3},
issn = {1936-7406},
url = {https://doi.org/10.1145/3482854},
doi = {10.1145/3482854},
journal = {ACM Trans. Reconfigurable Technol. Syst.},
month = {dec},
articleno = {23},
numpages = {32}
}

@article{goldstein2000piperench,
  title={PipeRench: A reconfigurable architecture and compiler},
  author={Goldstein, Seth Copen and Schmit, Herman and Budiu, Mihai and Cadambi, Srihari and Moe, Matthew and Taylor, R Reed},
  journal={Computer},
  volume={33},
  number={4},
  pages={70--77},
  year={2000},
  publisher={IEEE}
}

@inproceedings{chin2017cgra,
  title={CGRA-ME: A unified framework for CGRA modelling and exploration},
  author={Chin, S Alexander and Sakamoto, Noriaki and Rui, Allan and Zhao, Jim and Kim, Jin Hee and Hara-Azumi, Yuko and Anderson, Jason},
  booktitle={2017 IEEE 28th international conference on application-specific systems, architectures and processors (ASAP)},
  pages={184--189},
  year={2017},
  organization={IEEE}
}

@inproceedings{marshall1999reconfigurable,
  title={A reconfigurable arithmetic array for multimedia applications},
  author={Marshall, Alan and Stansfield, Tony and Kostarnov, Igor and Vuillemin, Jean and Hutchings, Brad},
  booktitle={Proceedings of the 1999 ACM/SIGDA seventh international symposium on Field programmable gate arrays},
  pages={135--143},
  year={1999}
}

@inproceedings{mirsky1996matrix,
  title={MATRIX: a reconfigurable computing architecture with configurable instruction distribution and deployable resources},
  author={Mirsky and DeHon},
  booktitle={1996 Proceedings IEEE Symposium on FPGAs for Custom Computing Machines},
  pages={157--166},
  year={1996},
  organization={IEEE}
}

@article{govindaraju2012dyser,
  title={Dyser: Unifying functionality and parallelism specialization for energy-efficient computing},
  author={Govindaraju, Venkatraman and Ho, Chen-Han and Nowatzki, Tony and Chhugani, Jatin and Satish, Nadathur and Sankaralingam, Karthikeyan and Kim, Changkyu},
  journal={IEEE Micro},
  volume={32},
  number={5},
  pages={38--51},
  year={2012},
  publisher={IEEE}
}

@article{singh2000morphosys,
  title={MorphoSys: an integrated reconfigurable system for data-parallel and computation-intensive applications},
  author={Singh, Hartej and Lee, Ming-Hau and Lu, Guangming and Kurdahi, Fadi J and Bagherzadeh, Nader and Chaves Filho, Eliseu M},
  journal={IEEE transactions on computers},
  volume={49},
  number={5},
  pages={465--481},
  year={2000},
  publisher={IEEE}
}

@inproceedings{cong2012charm,
  title={Charm: A composable heterogeneous accelerator-rich microprocessor},
  author={Cong, Jason and Ghodrat, Mohammad Ali and Gill, Michael and Grigorian, Beayna and Reinman, Glenn},
  booktitle={Proceedings of the 2012 ACM/IEEE international symposium on Low power electronics and design},
  pages={379--384},
  year={2012}
}

@inproceedings{cong2014fully,
  title={A fully pipelined and dynamically composable architecture of CGRA},
  author={Cong, Jason and Huang, Hui and Ma, Chiyuan and Xiao, Bingjun and Zhou, Peipei},
  booktitle={2014 IEEE 22nd Annual International Symposium on Field-Programmable Custom Computing Machines},
  pages={9--16},
  year={2014},
  organization={IEEE}
}

@inproceedings{zhang2018dnnbuilder,
  title={DNNBuilder: An automated tool for building high-performance DNN hardware accelerators for FPGAs},
  author={Zhang, Xiaofan and Wang, Junsong and Zhu, Chao and Lin, Yonghua and Xiong, Jinjun and Hwu, Wen-mei and Chen, Deming},
  booktitle={2018 IEEE/ACM International Conference on Computer-Aided Design (ICCAD)},
  pages={1--8},
  year={2018},
  organization={IEEE}
}

@ARTICLE{ahmed2024,
  author={Ahmad, Feroz and Zafar, Saima},
  journal={IEEE Embedded Systems Letters}, 
  title={SoC-Based Implementation of 1D Convolutional Neural Network for 3-Channel ECG Arrhythmia Classification via HLS4ML}, 
  year={2024},
  volume={},
  number={},
  pages={1-1},
  doi={10.1109/LES.2024.3354081}}

@inproceedings{wei2023fpga,
  title={FPGA-based microsecond-latency MHD mode tracking using high-speed cameras and deep learning on HBT-EP},
  author={Wei, Yumou and Arnold, David and Chandra, Rian and Dasilva, Nigel and Hansen, Christopher and Levesque, Jeffrey and Li, Boting and Notis, Matthew and Mauel, Michael and Navratil, Gerald and others},
  booktitle={APS Division of Plasma Physics Meeting Abstracts},
  volume={2023},
  pages={NP11--088},
  year={2023}
}

@INPROCEEDINGS{samsungCGRA,
  author={Kim, Changmoo and Chung, Mookyoung and Cho, Yeongon and Konijnenburg, Mario and Ryu, Soojung and Kim, Jeongwook},
  booktitle={2012 International Conference on Field-Programmable Technology}, 
  title={ULP-SRP: Ultra low power Samsung Reconfigurable Processor for biomedical applications}, 
  year={2012},
  volume={},
  number={},
  pages={329-334},
  doi={10.1109/FPT.2012.6412157}}

@misc{renesasSTP,
  author = {Renesas Electronics Corporation},
  title = {Renesas STP Engine (IP Core)},
  howpublished = {\url{https://web.archive.org/web/20231209041533/https://www.renesas.com/us/en/key-technologies/artificial-intelligence/stp-engine}},
  note = {Accessed: 2023-12-09}
}

@article{cgraSurvey,
author = {Liu, Leibo and Zhu, Jianfeng and Li, Zhaoshi and Lu, Yanan and Deng, Yangdong and Han, Jie and Yin, Shouyi and Wei, Shaojun},
title = {A Survey of Coarse-Grained Reconfigurable Architecture and Design: Taxonomy, Challenges, and Applications},
year = {2019},
issue_date = {November 2020},
publisher = {Association for Computing Machinery},
address = {New York, NY, USA},
volume = {52},
number = {6},
issn = {0360-0300},
url = {https://doi.org/10.1145/3357375},
doi = {10.1145/3357375},
journal = {ACM Comput. Surv.},
month = {oct},
articleno = {118},
numpages = {39}
}

@inproceedings{du2015shidiannao,
  title={ShiDianNao: Shifting vision processing closer to the sensor},
  author={Du, Zidong and Fasthuber, Robert and Chen, Tianshi and Ienne, Paolo and Li, Ling and Luo, Tao and Feng, Xiaobing and Chen, Yunji and Temam, Olivier},
  booktitle={Proceedings of the 42nd annual international symposium on computer architecture},
  pages={92--104},
  year={2015}
}

@inproceedings{umuroglu2017finn,
  title={Finn: A framework for fast, scalable binarized neural network inference},
  author={Umuroglu, Yaman and Fraser, Nicholas J and Gambardella, Giulio and Blott, Michaela and Leong, Philip and Jahre, Magnus and Vissers, Kees},
  booktitle={Proceedings of the 2017 ACM/SIGDA international symposium on field-programmable gate arrays},
  pages={65--74},
  year={2017}
}

@ARTICLE{zhang2023caffeine,
  author={Zhang, Chen and Sun, Guangyu and Fang, Zhenman and Zhou, Peipei and Pan, Peichen and Cong, Jason},
  journal={IEEE Transactions on Computer-Aided Design of Integrated Circuits and Systems}, 
  title={Caffeine: Toward Uniformed Representation and Acceleration for Deep Convolutional Neural Networks}, 
  year={2019},
  volume={38},
  number={11},
  pages={2072-2085},
  doi={10.1109/TCAD.2017.2785257}}

@ARTICLE{eyeriss,
  author={Chen, Yu-Hsin and Krishna, Tushar and Emer, Joel S. and Sze, Vivienne},
  journal={IEEE Journal of Solid-State Circuits}, 
  title={Eyeriss: An Energy-Efficient Reconfigurable Accelerator for Deep Convolutional Neural Networks}, 
  year={2017},
  volume={52},
  number={1},
  pages={127-138},
  doi={10.1109/JSSC.2016.2616357}}

@article{surveytaxo2022,
title = {A survey on hardware accelerators: Taxonomy, trends, challenges, and perspectives},
journal = {Journal of Systems Architecture},
volume = {129},
pages = {102561},
year = {2022},
issn = {1383-7621},
doi = {https://doi.org/10.1016/j.sysarc.2022.102561},
url = {https://www.sciencedirect.com/science/article/pii/S1383762122001138},
author = {Biagio Peccerillo and Mirco Mannino and Andrea Mondelli and Sandro Bartolini}
}

@article{surveycnndnn2023,
  title={From CNN to DNN Hardware Accelerators: A Survey on Design, Exploration, Simulation, and Frameworks},
  author={Juracy, Leonardo Rezende and Garibotti, Rafael and Moraes, Fernando Gehm and others},
  journal={Foundations and Trends{\textregistered} in Electronic Design Automation},
  volume={13},
  number={4},
  pages={270--344},
  year={2023},
  publisher={Now Publishers, Inc.}
}

@ARTICLE{surveyefficienthw2022,
  author={Dhilleswararao, Pudi and Boppu, Srinivas and Manikandan, M. Sabarimalai and Cenkeramaddi, Linga Reddy},
  journal={IEEE Access}, 
  title={Efficient Hardware Architectures for Accelerating Deep Neural Networks: Survey}, 
  year={2022},
  volume={10},
  number={},
  pages={131788-131828},
  doi={10.1109/ACCESS.2022.3229767}}

@article{vta,
  author       = {Thierry Moreau and
                  Tianqi Chen and
                  Ziheng Jiang and
                  Luis Ceze and
                  Carlos Guestrin and
                  Arvind Krishnamurthy},
  title        = {A Hardware-Software Blueprint for Flexible Deep Learning Specialization},
  journal      = {CoRR},
  volume       = {abs/1807.04188},
  year         = {2018},
  url          = {http://arxiv.org/abs/1807.04188},
  eprinttype    = {arXiv},
  eprint       = {1807.04188},
  bibsource    = {dblp computer science bibliography, https://dblp.org}
}

@article{leflow,
  author       = {Daniel H. Noronha and
                  Bahar Salehpour and
                  Steven J. E. Wilton},
  title        = {LeFlow: Enabling Flexible {FPGA} High-Level Synthesis of Tensorflow
                  Deep Neural Networks},
  journal      = {CoRR},
  volume       = {abs/1807.05317},
  year         = {2018},
  url          = {http://arxiv.org/abs/1807.05317},
  eprinttype    = {arXiv},
  eprint       = {1807.05317},
  bibsource    = {dblp computer science bibliography, https://dblp.org}
}

@article{di2021reconfigurable,
  title={A reconfigurable neural network ASIC for detector front-end data compression at the HL-LHC},
  author={Di Guglielmo, Giuseppe and Fahim, Farah and Herwig, Christian and Valentin, Manuel Blanco and Duarte, Javier and Gingu, Cristian and Harris, Philip and Hirschauer, James and Kwok, Martin and Loncar, Vladimir and others},
  journal={IEEE Transactions on Nuclear Science},
  volume={68},
  number={8},
  pages={2179--2186},
  year={2021},
  publisher={IEEE}
}

@article{duarte2018fast,
  title={Fast inference of deep neural networks in FPGAs for particle physics},
  author={Duarte, Javier and Han, Song and Harris, Philip and Jindariani, Sergo and Kreinar, Edward and Kreis, Benjamin and Ngadiuba, Jennifer and Pierini, Maurizio and Rivera, Ryan and Tran, Nhan and others},
  journal={Journal of instrumentation},
  volume={13},
  number={07},
  pages={P07027},
  year={2018},
  publisher={IOP Publishing}
}

@article{duarte2022fastml,
  title={FastML Science Benchmarks: Accelerating Real-Time Scientific Edge Machine Learning},
  author={Duarte, Javier and Tran, Nhan and Hawks, Ben and Herwig, Christian and Muhizi, Jules and Prakash, Shvetank and Reddi, Vijay Janapa},
  journal={arXiv preprint arXiv:2207.07958},
  year={2022}
}

@article{axani2024rfsoc,
  title={RFSoC-based front-end electronics for pulse detection},
  author={Axani, SN and Futagi, S and Garcia, M and Grant, C and Hosokawa, K and Ieki, S and Inoue, K and Ishidoshiro, K and Kawada, N and Matsumoto, Y and others},
  journal={Journal of Instrumentation},
  volume={19},
  number={03},
  pages={P03013},
  year={2024},
  publisher={IOP Publishing}
}

@techreport{arnold2023edge,
  title={Edge AI for accelerator controls (READS): beam loss deblending},
  author={Arnold, Jeremy and Austin, Mark and Berlioz, Jose and Bracey, Dave and Hanlet, Pierrick and Hazelwood, Kyle and Hu, Jerry Yao-Chieh and Ibrahim, Aisha and Jang, Jing and Liu, Han and others},
  year={2023},
  institution={Fermi National Accelerator Laboratory (FNAL), Batavia, IL (United States)}
}

@article{deiana2022applications,
  title={Applications and techniques for fast machine learning in science},
  author={Deiana, Allison McCarn and Tran, Nhan and Agar, Joshua and Blott, Michaela and Di Guglielmo, Giuseppe and Duarte, Javier and Harris, Philip and Hauck, Scott and Liu, Mia and Neubauer, Mark S and others},
  journal={Frontiers in big Data},
  volume={5},
  pages={787421},
  year={2022},
  publisher={Frontiers}
}

@inproceedings{resnet,
  title={Deep residual learning for image recognition},
  author={He, Kaiming and Zhang, Xiangyu and Ren, Shaoqing and Sun, Jian},
  booktitle={Proceedings of the IEEE conference on computer vision and pattern recognition},
  pages={770--778},
  year={2016}
}

@misc{brevitas,
  author       = {Alessandro Pappalardo},
  title        = {Xilinx/brevitas},
  year         = {2024},
  publisher    = {Zenodo},
  doi          = {10.5281/zenodo.3333552},
  url          = {https://doi.org/10.5281/zenodo.3333552}
}

@inproceedings{pointnet,
  title={Pointnet: Deep learning on point sets for 3d classification and segmentation},
  author={Qi, Charles R and Su, Hao and Mo, Kaichun and Guibas, Leonidas J},
  booktitle={Proceedings of the IEEE conference on computer vision and pattern recognition},
  pages={652--660},
  year={2017}
}

@article{borras2022open,
  title={Open-source FPGA-ML codesign for the MLPerf Tiny Benchmark},
  author={Borras, Hendrik and Di Guglielmo, Giuseppe and Duarte, Javier and Ghielmetti, Nicol{\`o} and Hawks, Ben and Hauck, Scott and Hsu, Shih-Chieh and Kastner, Ryan and Liang, Jason and Meza, Andres and others},
  journal={arXiv preprint arXiv:2206.11791},
  year={2022}
}

\end{document}